\documentclass[twoside, 12pt,a4paper]{article}
\usepackage[T1]{fontenc}
\usepackage{lmodern}
\usepackage{pgfplots,comment}

\usepackage{datetime}
\usepackage{aditya3}
\usepackage{enumitem}
\usepackage[]{titlesec}
\allowdisplaybreaks
\setlist{noitemsep,topsep=1pt}

\titlespacing*{\paragraph}{0pt}{0.5ex plus 1ex minus .2ex}{1em}
\tikzset{
    >=latex,
    pil/.style={
            draw,
      <-, 
      decorate,
      decoration={snake,,amplitude=.02cm, pre length=.2cm,post length=.2cm,}
              }}

\hypersetup{colorlinks=true,linkcolor=red, citecolor=blue, urlcolor=darkred}

\definecolor{BluishGreen}{RGB}{0,158,115}

\theoremstyle{exampstyle}
\newtheorem{proposition}{Proposition}

\usepackage[font=small]{caption}

\title{\bf \sc {\color{darkblue}the wrong kind of information}}

\author{ \begin{minipage}{0.3\textwidth}\centering 
Aditya Kuvalekar 
 \end{minipage}                  
 \begin{minipage}{0.3\textwidth}\centering 
 Jo\~ao Ramos  
 \end{minipage}                  
 \begin{minipage}{0.3\textwidth}\centering 
Johannes Schneider\footnote{Kuvalekar: University of Essex; Ramos: Queen Mary University and USC Marshall; Schneider: University of Mannheim \& Universidad Carlos III de Madrid. We are indebted to the editor, Nicola Persico, and three anonymous referees for excellent comments that improved the paper substantially. We thank Nageeb Ali, Rosella Argenziano, Heski Bar-Isaac, Dan Bernhardt, Dhruva Bhaskar, Antonio Cabrales, Odilon C\^amara, Marco Celentani, Joyee Deb, Siddharth Hari, Chad Kendall, Nenad Kos, Elliot Lipnowski, Antoine Loeper, Anthony Marino, John Matsusaka, Moritz Meyer-Ter-Vehn, Ignacio Ortu\~no, Harry Pei, Jacopo Perego, Maher Said, and Nico Schutz for helpful comments and discussions. Aditya Kuvalekar gratefully acknowledges support from  the Ministerio Economia y competitividad grant PGC2018-09159-B-I00. Johannes Schneider gratefully acknowledges financial support from the German Research Foundation (DFG) through CRC TR 224 (Project B03), Agencia Estatal de Investigación (PID2019-111095RB-I00 and PID2020-118022GB-I00), Ministerio Economía y Competitividad (ECO2017-87769-P), and Comunidad de Madrid (MAD-ECON-POL-CM H2019/HUM-5891 and EPUC3M11 (V PRICIT)).}  
 \end{minipage}                  
 }

\usetikzlibrary{external}

\date{\monthname, \the\year} 

\begin{document}

\maketitle
\begin{abstract}
Agents, some with a bias, decide between undertaking a risky project and a safe alternative based on information about the project's efficiency. Only a part of that information is verifiable. Unbiased agents want to undertake only efficient projects, while biased agents want to undertake any project. If the project causes harm, a court examines the verifiable information, forms a belief about the agent's type, and decides the punishment. Tension arises between deterring inefficient projects and a chilling effect on using the unverifiable information. Improving the unverifiable information always increases overall efficiency, but improving the verifiable information may reduce efficiency.

\end{abstract}

\onehalfspacing

\section{Introduction}
From politicians to doctors to civil servants, examples abound of people avoiding risky but socially efficient decisions for fear of being sued. When faced with a risky decision, individuals rely on information to evaluate their action's costs and benefits. However, if the action results in perverse consequences and litigation ensues, only part of that information is verifiable by a court. For example, policy makers decide on reforms on the basis of experts' reports, which are verifiable by outsiders, but also on their own expertise; doctors decide the course of treatment for a patient based on test results and also their examinations. If the policy implemented or the treatment prescribed fails, these agents face the threat of punishment. Anticipating that threat, well-intentioned agents tend to overweight the verifiable information and ignore useful yet unverifiable information. An agent fears that, in litigation, the court---relying only on the verifiable portion of the information---will mistakenly decide that he acted recklessly and for personal benefit rather than in society's interest. In consequence, a well-intentioned agent suffers from a \emph{chilling effect:} He shies away from a socially efficient action for the fear of being sued.

Lawmakers, in turn, may be motivated to deter  biased agents---agents whose preferences differ from the lawmakers'---from undertaking inefficient projects. But they must also consider how the threat of punishment affects agents' use of information. They must strike a balance between deterring biased agents from taking socially inefficient actions and encouraging unbiased agents to use both the verifiable and the unverifiable information. Therefore, the optimal design of the law depends on the precision of both the verifiable and the unverifiable information. This dependency raises the questions that motivate our paper: How does the precision of information---verifiable and unverifiable---affect the quality of agents' decisions? Can better information exacerbate the chilling effect? And, if so, can the lawmaker design the law so that the benefits of the superior information outweigh the costs of the chilling effect? Our paper studies and  answers these questions. We show that in equilibrium agents may make less efficient decisions if the verifiable information becomes more precise; but the efficiency of agents’ decisions always improves with greater precision of the unverifiable information. 

To capture the above-described environment, we employ the following simple model. There are three players: a designer of the law, a court, and an agent. The agent chooses between undertaking a risky project and taking a safe alternative. The designer wants the agent to undertake the risky project if and only if it is likely to succeed; otherwise the designer prefers the safe alternative. The agent, in turn, can be unbiased or biased toward taking the risky project. The court serves as an exogenous institution that applies the law with the goal of screening out and punishing biased agents. The designer moves first and determines the maximum punishment the court can impose on the agent. Then the agent decides whether to undertake the risky project or to take the safe alternative. To assess the likelihood of success, he relies on two pieces of information about the risky project---one verifiable and one unverifiable. If the agent undertakes the project and it fails, he is taken to court. The court examines the verifiable information and forms a belief about whether the agent is biased. Based on this belief and the limits set by the designer, it decides whether and how much to punish the agent. 

To make an efficient decision, an unbiased agent uses both the verifiable and the unverifiable information. At times, the agent should undertake the project even when the verifiable information favors the safe action. A sufficiently informative unverifiable signal favoring the risky project may trump the negative verifiable information. This observation leads to the key tradeoff for the designer: On the one hand, if the threat of punishment is large in case of a failure, the unbiased agent fears that he will be convicted by the court. This induces the chilling effect: the agent ignores the unverifiable information and bases his decision primarily on the verifiable information. On the other hand, if the threat of punishment is low, it will fail to deter the biased agent from undertaking the risky project even when it is inefficient. Our main result shows that the cost of deterrence---the chilling effect---becomes larger as the verifiable information becomes more precise. In fact, it can overpower the benefits of improved information and lead to a reduction in ex ante efficiency (Proposition \ref{prop:main_result}). In contrast, the cost of deterrence becomes smaller as the unverifiable information becomes more precise, leading to an unambiguous increase in ex ante efficiency (Proposition \ref{prop:pyincreasegood}).

The driving force behind our main results is the difference between the two agents' considerations when they contemplate  undertaking the project or taking the safe action. To illustrate the mechanism, assume that the agents get punished if the risky project is implemented, it fails, and the verifiable information favors the safe option. The larger the agents' punishment, the more the safe alternative appeals to both agents. With more precise verifiable information, the project has a higher chance of failure---and hence punishment---when the verifiable information favors the safe alternative. In other words, the fear of punishment increases for both types with more precise verifiable information. In addition to that punishment effect, there is a second effect relevant only for the unbiased type. Since the unbiased agent wishes to undertake the project only when it is likely to succeed, when a more precise verifiable signal favors the safe alternative, the risky project is even less attractive than before.  Therefore, there is a stronger chilling effect on the unbiased type: he is more afraid to undertake the project even when it is efficient if that requires going against his verifiable information. As a consequence, deterrence is now accompanied by a stronger chilling effect and, as Proposition \ref{prop:main_result} shows, efficiency decreases.

The same argument does not hold when the unverifiable information becomes more precise. In this case, deterrence becomes easier.  The reason is that it is socially optimal to deter the biased agent from undertaking the project when both the verifiable and the unverifiable information favor not choosing it. More precise unverifiable information leads to a stronger punishment effect here but on the biased type only. The unbiased type would never wish to undertake the project in such situations even without punishment. At the same time, the unbiased agent is now more optimistic about the project's success whenever the unverifiable information recommends undertaking it. Thus, he is more likely to make use of the unverifiable information; the chilling effect declines. Taking both effects together, screening gets easier for the designer and, as Proposition \ref{prop:pyincreasegood} shows, efficiency increases. 

 Our results provide a cautionary tale in a world in which we observe constant improvements in  available information, both verifiable and unverifiable. For example, doctors get better diagnostic tools; politicians get access to more specialized expert reports; civil servants are expected to use newer software to compare prices. While marginal improvements in unverifiable information are always welfare improving, one should be careful when marginally improving the precision of verifiable information. That is, improvements in technologies such as better diagnostic tools for doctors may simply not substitute for similar improvements acquired through experience. Even worse, such improvements could backfire. 

Finally, our results extend to various alternative institutional settings. In Section \ref{sec:Robustness} we discuss a large range of model generalizations to show how the arguments translate to other environments. The key model ingredients driving  our result are the following: (i) some but not all agents strive for efficiency; (ii) it is common knowledge that agents possess valuable information beyond what is ex post verifiable; (iii) agents are screened ex post based on outcomes and the verifiable information. All three elements are present in several real-world settings beyond  the legal system. In Section \ref{sub:examples} we discuss a variety of settings with and without formal courts to highlight the applicability of our arguments. 

\subsection{Related Literature}

At a superficial level, the main takeaway of our paper---that superior verifiable information may reduce welfare---is reminiscent of several   literatures. While our results are connected to some of them, we highlight in this section the  different economic forces that lead to our results. To this end, we discuss each of the  literatures separately.

\paragraph{Exclusion of Verifiable Information.} Federal Rules of Evidence 403 and 404 allow  judges to exclude evidence with probative value. \citet{lester2012information} argue that such exclusion may increase welfare. A cost-minimizing fact finder may opt to evaluate evidence with lower statistical power, as it is less costly to do so. \citet{BullWatson219} provide a model of ``robust litigation,'' in which litigants can choose whether to present hard, verifiable information. They show that, depending on the strength of the litigant's private signal relative to that of the hard information, the hard information can be misleading and lead to a loss in welfare. 

Unlike \citet{BullWatson219}, we abstract from any signaling concerns in the disclosure of the hard information and focus on a setting in which disclosure is mechanical. In this setting, inefficiency is caused by the agent's hesitation to take an efficient action due to the chance that such an action will lead to (i) harm and (ii) punishment. 

In our setting, the defendant's action is publicly observed, and the court's role is to determine whether the intent behind the action was suspect.  In line with \citet{10.2307/1123746} and \citet{schrag1994crime}, we are interested in how evidence shapes agents' behavior  outside the courtroom (and thus how it affects society's welfare). \citet{10.2307/1123746} and \citet{schrag1994crime} study how the law can deter an agent whose preferences do not align with society's. We complement this setting by introducing an unintended side effect of deterrence: the chilling effect on an unbiased agent.\footnote{All four papers discuss their findings in light of Federal Rule 404, which concerns the exclusion or inclusion of character evidence in the trial. In our environment, all evidence of the agent's character comes from his behavior and not from  observable character traits. Thus, the court in our model complies with Federal Rule 404.}  

\paragraph{The Chilling Effect.} The chilling effect has been recognized in the literature. An early attempt to capture it formally is in \citet{garoupa1999economics}. In more recent work, \citet{kaplow2011optimal,kaplow2017optimal,kaplow2018optimal} documents the need to balance deterrence against the chilling effect in a variety of settings. 

We build on this literature by taking the chilling effect as the starting point of our analysis. Allowing the punishment scheme to vary with the quality of information, we explore whether the chilling effect can be mitigated through the combination of superior information and an optimal  judicial system. 

\paragraph{Other Side Effects of Deterrence.} A small literature has considered other, orthogonal side effects of deterrence. \Citet{stigler1970optimum} argues that imposing a harsh punishment for minor crimes may erode societies' willingness to punish any crime and suggests intermediate punishment as a remedy. \citet{10.1111/1467-937X.00162} points out that democratic societies have strategic reasons to limit punishment since an erroneous interpretation of the law by courts may hurt the ``wrong'' part of the population. \citet{pei2019crime} show that  severe punishment reduces the number of crimes that witnesses report, thereby reducing the cost of committing a crime. Intermediate punishments can deter some individuals from committing crimes, but those that commit  crime are likely to commit several crimes. Unlike these researchers, we concentrate on how the quality of information affects the tradeoff between deterrence and the chilling effect.

\paragraph{Incorporating Different Types of Information.} 
Our main comparative static---increasing the precision of verifiable information can harm welfare\textemdash is reminiscent of \citet{morris2002social} if one views verifiable (unverifiable) information as public (private) information. However, our channel differs from theirs. Coordination motives\textemdash the main driver in \citet{morris2002social}\textemdash are entirely absent in our model. To highlight the difference between the environments, consider the setting in which the private information is very precise. Because of coordination motives, small increases in the precision of public information can harm welfare in the setting of \citet{morris2002social}. Players overweigh public information, leading to a welfare reduction if it is sufficiently noisy. An analogous result does not appear in our setting. With precise private information, agents can be screened effectively via the outcome.

In a principal-agent setting, \cite{vidal2007should} study a problem in which the principal has two pieces of information, only one of which can be shared with the agent before he chooses his effort level. They show that sharing information may harm welfare. While our setting is outwardly similar to theirs, in ours  the agent (not the principal) possesses the information and the punishment is (endogenously) determined ex ante to optimally discipline the agent. The economic forces in our environment do not rely on an agent that is suspicious of the selection of the signal received but on an agent afraid of being punished for relying on all of his available information.

\paragraph{Contract Theory.} The closest papers in the contracting literature are \citet{prendergast1993yes} and  \citet{prat2005wrong}. In a principal-agent setting, \citet{prendergast1993yes} focuses on how to incentivize an agent to acquire relevant information at a cost. He highlights how the agent may focus more on acquiring information about the principal's prior belief  than about the underlying state of the world.

\citet{prat2005wrong} argues that the \emph{content of information} leads to qualitatively different effects of increased precision. While information about consequences is beneficial, that about actions is harmful. 
We view our exercise as complementary to \citeauthor{prat2005wrong}'s (\citeyear{prat2005wrong}) and \citeauthor{prendergast1993yes}'s (\citeyear{prendergast1993yes}).  While they focus  on situations in which the underlying information is about different objects (state of the world versus the principal's prior belief in \cite{prendergast1993yes} and consequences versus actions in \cite{prat2005wrong}), both signals in our framework provide information about the same object. We show how the \emph{nature of the information} about the same object\textemdash the quality of the project\textemdash affects welfare. 

\section{Model}

\begin{figure}\centering 
  \begin{tikzpicture}[]
\draw (0,0) -- (12,0);
\draw (0,-.1) -- (0,.1) node[above, align=center]{designer \\ selects\\ $\overline{F}>0$};
\draw (2,-.1)node[below, align=center]{nature \\ draws \\ $\{\theta,\omega,x,y\}$} -- (2,.1); 
\draw (4,-.1) -- (4,.1) node[above, align=center]{agent \\ learns \\ $(\omega,x,y)$};
\draw (6,-.1) -- (6,.1) node[above, align=center]{agent \\ decides\\ $a{ \in }\{0,1\}$};
\draw (8,-.1) -- (8,.1) node[above, align=center]{court \\ observes \\ $(a,\theta,x)$};
\draw (10,-.1) -- (10,.1) node[above, align=center]{court\\ decides  \\$F(x) {\in }[0,\overline{F}]$};
\draw (12,-.1) -- (12,.1) node[right, align=center]{game \\ ends};
\node at (7,0) [below]{$a{=}1$};
\node at (9,0) [below]{$\theta{=}${-}$1$};
\draw[-|] (6,0)--(6,-1.5) node[below, align=center]{game\\ends};
\draw[-|] (8,0)--(8,-1.5) node[below, align=center]{game\\ends};
\node at (6,-1.1) [right]{$a{=}0$};
\node at (8,-1.1) [right]{$\theta{=}1$};
\end{tikzpicture}
  \caption{\emph{Timing of the game.} The designer selects the maximum punishment, $\overline{F}$. The agent observes his type realization, $\omega$, and the realization of the two noisy signals, $(x,y)$, about the risky project's quality. Based on $(\omega,x,y)$, the agent decides whether to take the risky action, $a{=}1$, or the safe action, $a{=}0$. If the agent takes the risky action, the court observes the realized project quality, $\theta$, and the realization of the verifiable signal, $x$. If the project fails ($\theta{=-}1$), the court selects a punishment $F(x)\in[0,\overline{F}]$. Then payoffs realize.}\label{fig:timeline}
\end{figure}
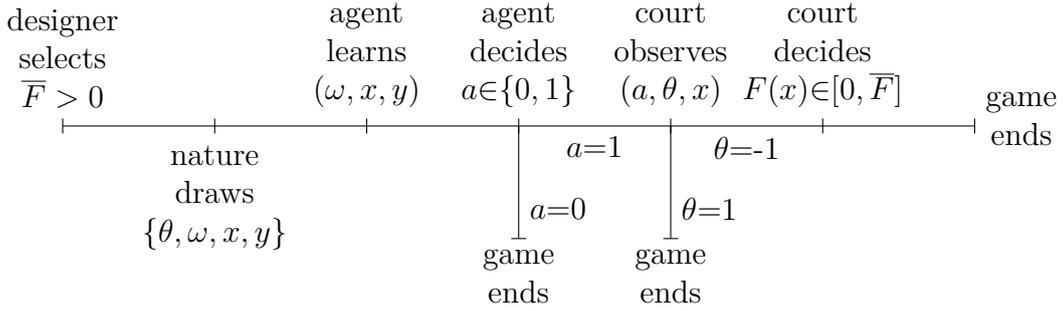

There are three players: an agent (``he''), a court (``it''), and a designer (``she''). The agent decides whether to undertake a risky project that may succeed or fail. The agent is uncertain about the quality of the project and relies on the information available to him when making a decision. If the agent undertakes the risky project, and it  fails, the court examines the verifiable part of the agent's information and decides the punishment. The court applies the law, which depends on the designer's initial choice of the maximum punishment. Figure \ref{fig:timeline} summarizes the basic model structure.

\paragraph{Project Quality and Information.} The project's quality is either good ($\theta = 1$) or bad $(\theta = -1$). If undertaken, a good project succeeds and a bad project fails. 

The ex ante probability that the project is good is $\beta$. There are two imperfectly informative signals about the project's quality: the verifiable information and the unverifiable information. The verifiable information is a random variable $\pubsigr$ with realization $x \in \{-1,1\}$. The precision of the verifiable information is given by $p_x:=\prob(\pubsigr=\theta) \in (1/2,1)$, the probability that the verifiable signal matches the quality of the project. Analogously, the unverifiable information is a random variable $\pvtsigr$ with realization $y \in \{-1,1\}$ and precision $p_y:=\prob(\pvtsigr=\theta)\in (1/2,1)$. We summarize the informational environment by $S:=(\beta,p_x,p_y)$. The signals $\pubsigr$ and $\pvtsigr$ are independent conditional on the project's state $\theta$.\footnote{To keep the analysis simple, our baseline signal structure is very stylized. Continuous signals are discussed in Appendix \ref{sec:GeneralStructure}; in Appendix \ref{sub:asy} we discuss state-dependent levels of precision; and in Appendix \ref{sub:CondDep}, we cover conditionally dependent signals.}

\paragraph{\textbf{Designer}.} At the beginning of the game, the designer chooses the maximum punishment, $\overline{F}$, the court can inflict on the agent. The designer receives a payoff of 1 from a successful project and a payoff of -1 from a failed project. If no project is undertaken, she receives a payoff of 0.

\paragraph{Agent.} The agent is privately informed about his type $\omega$. He can be unbiased ($\omega=u$) or biased ($\omega=b$).  The common prior $\gamma$ denotes the ex ante probability that $\omega = u$. The agent observes the realizations $x$ and $y$ of the two signals and decides whether to act ($a=1$)---that is, undertake the project---or not ($a=0$).

The ex post payoffs, $u^\omega$, of an agent of type $\omega$ from his action are given by

\[u^u(a,\theta)= a \theta \quad \text{ and } u^b(a,\theta) = a.\] 

An unbiased agent benefits from successful projects but suffers from failed projects; a biased agent benefits whenever he acts. In addition, the court can reduce an agent's utility by punishment $F$.

\paragraph{Court.} The court observes the agent's action  $a$ and the realization of the verifiable information $x$. It has no access to the unverifiable information. Based on the information, the court applies the law and potentially inflicts punishment $F\in [0,\overline{F}]$. We assume the following on the court's behavior: (i) The court can only punish upon harm, that is, if the risky project fails,\footnote{This assumption is motivated by realism and does not affect our results. We show this in Section \ref{Section: conviction upon a=0} which also provides further discussion on this point.} and  (ii) the court is set to screen agents. That is the court receives a positive payoff $F$ if it inflicts $F$ on a biased agent. It suffers a loss $FL$ if it does so on an unbiased agent. $L>0$ is a scaling parameter.

\paragraph{Welfare.} 
Let $a^\omega(x,y)$ be the type-$\omega$ agent's probability of acting on $(x,y)$, and let $F(x)$ be the court's punishment strategy. 
We define welfare, $W(\cdot)$, to be the ex-ante expected utility of the designer. Formally, 
\begin{align*}
W(a^u(\cdot), a^b(\cdot), F(\cdot), \of; S,\g) := \E_{\omega,x,y,\theta}[ a^\omega (x,y) \theta]
\end{align*}
We focus on the designer-optimal perfect Bayesian equilibria.

\paragraph{On The Court's Objective Function.} Before moving to the analysis, we pause to discuss our assumptions about the court's behavior. Together, they capture the doctrine  ``actus reus non facit reum nisi mens sit rea'' (the act is not culpable unless the mind is guilty). The doctrine requires that a person can be found guilty only if there has been (a) a physical element---an unlawful action---and (b) a mental element---a violation of the standards of care such as negligence or an intention to harm. 

In our model, taking a risky action that  fails serves as the physical element necessary for conviction. Regarding the mental element, a biased agent intrinsically exercises a lower standard of care as compared to the unbiased agent. Thus, his bias constitutes the guilty mind. Taken together, actus reus and mens rea imply the courts' objective: it wishes to convict the biased agent for undertaking the risky project that fails.

Moreover, we have chosen the concept of \emph{subjective} mens rea in our baseline setting. The guilty mind depends on the inferred preferences of the agent; the law aims to screen agents' types. An alternative interpretation is \emph{objective} mens rea. In that case, the guilty mind depends on the inferred unverifiable information of the agent and the time of decision making; the law aims to screen the agent's information set. 

Historically, at least criminal law often relies on subjective mens rea. In tort law cases, mens rea plays a less formal role. However, informally the defendant's (perceived) type remains important in establishing liability \citep[see, for example,][for a discussion]{cane2000mens}. While both types of mens rea provide similar results, subjective mens rea appears more appropriate for two reasons. It provides a sharper description of our central tradeoff, and a welfare-maximizing designer prefers it over objective mens rea. We provide further discussion and examples of the settings we have in mind in Section \ref{Subsection: mens rea different interpretation}.

\section{Analysis} 
\label{sec:analysis}
We characterize the equilibria  using backward induction. We first analyze the court's best response, then that of the agent. Finally, we determine the designer's optimal choice. In Section \ref{sub:comparative_statics}, we present our main result: increasing  the precision of the unverifiable information, $p_y$, always improves welfare. In contrast, increasing  the precision of the verifiable information, $p_x$, may reduce welfare.

\subsection{Court's Best Response}\label{sub:cout_s_best_response}
After observing the agent's action, the realization of the state, and the realization of the verifiable signal, the court decides how much to punish the agent. It takes the agent's equilibrium behavior and the maximum punishment set by the designer, $\overline{F}$, as given. By assumption the court can only convict if the project failed---that is, $a=1$ and $\theta=-1$. Recall that the court receives a payoff of $F$ if it convicts a biased agent  and takes a loss of $FL$ if it convicts an unbiased agent. No conviction implies $0$ payoff. Thus, when deciding on the punishment, the court uses Bayes' rule to form a belief, denoted by $\gamma_x$, about the probability that the agent is unbiased. In calculating $\gamma_x$ it takes into account both the verifiable information and the agent's equilibrium behavior. Subject to that belief, the court's expected payoff from conviction is

\[(1-\gamma_x)F-\gamma_xFL.\]

The court chooses to convict the agent only if the above is weakly larger than zero---the payoff of not convicting. Because the interim payoff of convicting is decreasing on $\gamma_x$,  there is a unique belief that makes the court indifferent between convicting and not: $\overline{\gamma}:=1/(1+L)$. The court's optimal strategy is a simple cutoff strategy: it inflicts the maximum punishment, $F=\overline{F}$, if $\gamma_x < \overline{\gamma}:=1/(1+L)$ and no punishment, $F=0$, if $\gamma_x > \overline{\gamma}$. The court is indifferent between  sentences if $\gamma_x = \overline{\gamma}$. The relevant case for our purposes is $\gamma>\overline{\gamma}$ which we assume from now on.

\subsection{Agent's Best Response} 
\label{sub:agent_s_best_response}

The agent observes $(x,y)$ and decides whether to act. If he decides not to act, he receives a payoff of $0$. The payoff from acting depends on whether the project ultimately succeeds or fails and on the punishment the court inflicts if it fails. The agent uses the information $(x,y)$ to update his prior belief via Bayes' rule. He forms a posterior belief, $\beta_{xy}$, that describes the interim probability that the project is good. Taking the court's decision as given, his interim expected payoff from acting is as follows:
\[\beta_{xy} u^\omega(a=1,\theta=1)+ (1-\beta_{xy}) (u^\omega(a=1,\theta=-1) -  F(x))\]

The agent prefers acting only if the above is larger than $0$---the payoff from not acting. Because the interim expected payoff is monotonically increasing in $\beta_{xy}$, the agent follows a cutoff strategy with type-specific cutoffs

\begin{align}\label{Equation: cutoff beliefs}
\overline{\beta}^u(F(x)) := \frac{F(x)+1}{F(x)+2} \quad \text{ and } \quad \overline{\beta}^b( F(x)) := \frac{ F(x)-1}{F(x)}.
\end{align}

A type-$\omega$ agent strictly prefers to act if $\beta_{xy}> \overline{\beta}^\omega(F(x))$, prefers to not act if $\beta_{xy}< \overline{\beta}^\omega( F(x))$, and is indifferent between the two if $\beta_{xy}=\overline{\beta}^\omega(F(x))$. Notice that $\overline{\beta}^u(F(x)) > \overline{\beta}^b(F(x))$. Therefore, whenever the unbiased agent weakly prefers acting, the biased agent strictly prefers to act. 

We abuse notation slightly and denote by $a^\omega(x,y)$ the probability that an agent of type $\omega$ acts, taking $F(x)$ and $\overline{F}$ as given. 

\subsection{Designer's Best Response} 
\label{sub:optimal_penal_code}
The designer selects the maximum punishment, $\overline{F}$, with the goal of maximizing welfare.
Notice that, if $F(x) = 0$,  an unbiased agent would act on $(x,y)$ whenever $\b_{xy}> \frac12$. Since the unbiased agent's preferences coincide with those of the society's, his actions too would coincide with the society's preferred actions when $F(x) = 0$. Therefore, we say that it is interim efficient to act on $(x,y)$ if $\b_{xy}> \frac12$. 

In the main text, we focus on  environments $S$ such that \[\beta_{xy}\geq 1/2 \Leftrightarrow \max\{x,y\}=1.\]
That is, we consider the cases in which it is (interim) efficient to act if and only if the agent receives at least one positive signal. 
We chose this case because it illustrates our main point most clearly.\footnote{Our results are not specific to this case. 
For a formal treatment, see Appendix \ref{sec:online_appendix}.}

\paragraph{Basic Tradeoff.} The two equations in \eqref{Equation: cutoff beliefs} highlight the main tradeoff in designing the maximum punishment level $\overline{F}$. If the expected punishment, $F(x)$, is too low, the biased agent acts even if it is inefficient to do so. If $F(x)$ is too high, the unbiased agent suffers from the \emph{chilling effect}: the fear of being punished if the project fails results in not acting when it is efficient to act.

The optimal punishment scheme balances the deterrence of the biased agent with the encouragement of the unbiased agent.

\subsection{Optimal Punishment Scheme}

Having laid out the agent's and the court's incentives, we now proceed to solve for the optimal punishment scheme the designer will set, given the information structure. 

First, notice that it is efficient to act when the verifiable information is positive, $x=1$, regardless of the realization $y$. Indeed, if both agents act on $x=1$, the court's posterior probability is equal to the prior $\gamma$. Since $\g > \overline{\gamma}$, the court never punishes an acting agent when $x=1$. 
Conflict arises only when the verifiable information is negative, $x=-1$. Here it is efficient to act on positive unverifiable information, $y=1$, but efficient to not act on negative unverifiable information, $y=-1$.

Given $x=-1$, the designer wishes to deter the biased agent from acting when $y=-1$ while incentivizing the unbiased agent to act when $y=1$. This leads to the following two natural questions that guide our analysis. Assuming that the agent gets punished when the project fails and $x=-1$, we ask: 
\begin{enumerate}
\item What is the minimum punishment, $F^b$, that prevents the biased type from acting when receiving  negative unverifiable information, $y=-1$? 
\item What is the maximum punishment, $F^u$, that will allow the unbiased type to act when receiving positive unverifiable information, $y=1$?
\end{enumerate}
Invoking the agent's best response, we obtain
\begin{equation}\label{Equation:Fd and Fh}
F^u = \frac{2\b_{-1,1}-1}{1-\b_{-1,1}} \quad \text{ and } \quad  F^b = \frac{1}{1-\b_{-1,-1}}.
\end{equation}
Whether $F^u > F^b$ or $F^b > F^u$ depends on the information structure, $S$. As we shall see, the optimal equilibrium has a different structure depending on whether $F^u > F^b$ or vice versa.

To understand the difference between these two cases, it is helpful to consider them separately. To illustrate the underlying intuition, we ignore the court's incentives momentarily and assume that the court  punishes with $F(-1)=\overline{F}$. Then, by the definitions of $F^b$ and $F^u$, an unbiased agent strictly prefers to act on $y=1$ if $\of < F^u$, while the biased agent strictly prefers to act on $y=-1$ if $\of < F^b$.

\paragraph{Case $\mathbf{F^b > F^u}$.} If $\of \le F^u$, then the unbiased agent acts on $y=1$ while the biased agent acts on all signal realizations. The punishment is too low to achieve any deterrence. Therefore, welfare is constant for all $\of \in [0,F^u]$, and it is without loss to assume that the designer sets $\of = 0$; she offers the agent a \emph{universal free pass} (Table \ref{Table:Fh less than Fd and Fbar =0}). If $\of \in (F^u, F^b)$, then the unbiased agent will prefer to not act on $y=1$, yet the biased agent will continue to act on $y=-1$. But then the designer can attain  strictly higher welfare by setting $\of = F^b$: doing so deters the biased agent from acting on $y=-1$, leaving the unbiased agent's behavior unchanged as seen in Table \ref{Table:Fh less than Fd and Fbar =Fd}. Welfare improves. 

Hence, the optimal punishment scheme is either a universal free pass ($\overline{F}=0$) or $\overline{F}=F^b$. The latter deters the  biased agent from acting on two negative signals at the cost of fully chilling the unbiased agent's action on $y=1$. With $\overline{F}=0$, the court's incentives play no role, while with $\overline{F} = F^b$, the court expects only the biased agent to act on $x=-1$. It is  optimal to set $F(-1)=\overline{F}$ given the agent's behavior.

\paragraph{Case $\mathbf{F^u > F^b}$.} Here it is possible to partially deter the biased agent without imposing a chilling effect on the unbiased agent. By setting $\overline{F}=F^b$, the biased agent is deterred from acting on $y=-1$, whereas the unbiased agent is encouraged to act on $y=1$. Notice, however, that in this case, the biased agent cannot be fully deterred from acting on $y=-1$ in equilibrium. The reason comes from the court's equilibrium behavior, an issue we have so far ignored. If the biased agent does not act on $y=-1$, and if both types of agents act on $y=1$, then the court's posterior belief about the agent's type upon seeing a failure and when $x=-1$ is $\g > \overline{\gamma}$. The court will not convict the agent. Naturally, the biased agent could exploit the court's behavior and act on $y=-1$. Therefore, in the welfare-maximizing equilibrium, the court must be indifferent about convicting the agent. To make the court indifferent, the biased agent must mix with an interior probability.\footnote{In other equilibria,  the biased agent acts with high probability on $y=-1$ and the court strictly prefers conviction. However, these equilibria are not welfare maximizing.} 
This equilibrium is summarized in Table \ref{Table:Fh greater than Fd and Fbar =Fd}. Alternatively, we can have $\of = F^u$, which fully deters the biased agent, $a^b(-1,-1)=0$, but at the cost of a partial chilling effect, $a^u(-1,1) < 1$, as seen in Table \ref{Table:Fh greater than Fd and Fbar =Fh}. Again, the reason for the mixing of the unbiased type on $(-1,1)$ is to provide incentives to the court for conviction upon failure and $x=-1$.

\begin{table}[tb]
\caption{Strategy profiles in the optimal equilibria}
\label{Table: equilibrium strategy profiles}
\begin{subtable}[t]{0.95\linewidth}
    \caption*{When $F^b > F^u$}
    \label{Table: Fd greater than Fh}
    \centering
\begin{subtable}[t]{0.45\linewidth}
      \caption{When $\of = 0$}
              \label{Table:Fh less than Fd and Fbar =0}
      \centering
         \begin{tabular}{lll}
           $(x,y)$ &  $a^u$ & $a^b$\\
           \hline
           (-1,-1) & 0 & 1 \\
           (-1,1) & 1 & 1
        \end{tabular}
\end{subtable}
\begin{subtable}[t]{0.45\linewidth}
      \caption{When $\of = F^b$}
              \label{Table:Fh less than Fd and Fbar =Fd}
      \centering
         \begin{tabular}{lll}
           $(x,y)$ &  $a^u$ & $a^b$\\
           \hline
           (-1,-1) & 0 & 0 \\
           (-1,1) & 0 & 1
        \end{tabular}
\end{subtable}
\end{subtable}
\bigskip
\begin{subtable}[t]{0.95\linewidth}
    \caption*{When $F^u > F^b$}
    \label{Table: Fd less than Fh}
    \centering
\begin{subtable}[t]{0.45\linewidth}
       \caption{When $\of = F^b$}
              \label{Table:Fh greater than Fd and Fbar =Fd}
      \centering
         \begin{tabular}{lll}
           $(x,y)$ &  $a^u$ & $a^b$\\
           \hline
           (-1,-1) & 0 & $\eta^b$ \\
           (-1,1) & 1 & 1
        \end{tabular}
\end{subtable}
\begin{subtable}[t]{0.45\linewidth}
     \caption{When $\of = F^u$}
              \label{Table:Fh greater than Fd and Fbar =Fh}
      \centering
         \begin{tabular}{lll}
           $(x,y)$ &  $a^u$ & $a^b$\\
           \hline
           (-1,-1) & 0 & 0 \\
           (-1,1) & $\eta^u$ & 1
        \end{tabular}
\end{subtable}
\end{subtable}
\end{table}

As we have seen, the ranking of the critical levels $F^u$ and $F^b$ determines how deterrence and the chilling effect pair. For example, if $F^b>F^u$, full deterrence also implies a maximal chilling effect. If $F^u>F^b$, full deterrence is possible at a lower cost. The ranking depends on the information structure $S$ and, in particular, on the levels of precision for the verifiable and the unverifiable information, $p_x$ and $p_y$. Lemma \ref{lem:FdminusFhincreasing} characterizes the effect of a change in $p_x$ and $p_y$ on the difference $F^b-F^u$. It is at the heart of our main result.

\begin{lemma}\label{lem:FdminusFhincreasing}
The difference between the critical punishment levels, $F^b-F^u$, is continuous in both precision levels, $p_x$ and $p_y$. The difference is \emph{increasing} in $p_x$ and \emph{decreasing} in $p_y$.
\end{lemma}

To understand why the difference is increasing in  precision $p_x$, first note that both $F^b$ and $F^u$ are decreasing in $p_x$. The likelihood of failing, conditional on $x=-1$, increases with $p_x$, and thus the agent expects\textemdash ceteris paribus\textemdash a higher punishment. However, the increases in both punishment and probability of failure affect the different types in different ways. Because of the misalignment of preferences between the unbiased and the biased agent, $F^u$ falls faster than $F^b$. The biased agent suffers only indirectly from the higher failure rate\textemdash through the higher punishment (the punishment effect). The unbiased agent also suffers  directly\textemdash through the failure itself (the outcome effect). 

The reason that the difference is decreasing in $p_y$ is more direct. Following an increase of $p_y$, both outcome and punishment effects encourage the unbiased agent to act on $y=1$; thus, $F^u$ increases. The punishment effect discourages the biased agent from acting on $y=-1$; thus, $F^b$ decreases.
\begin{figure}[h!]
\centering
\includegraphics[width=0.7\textwidth]{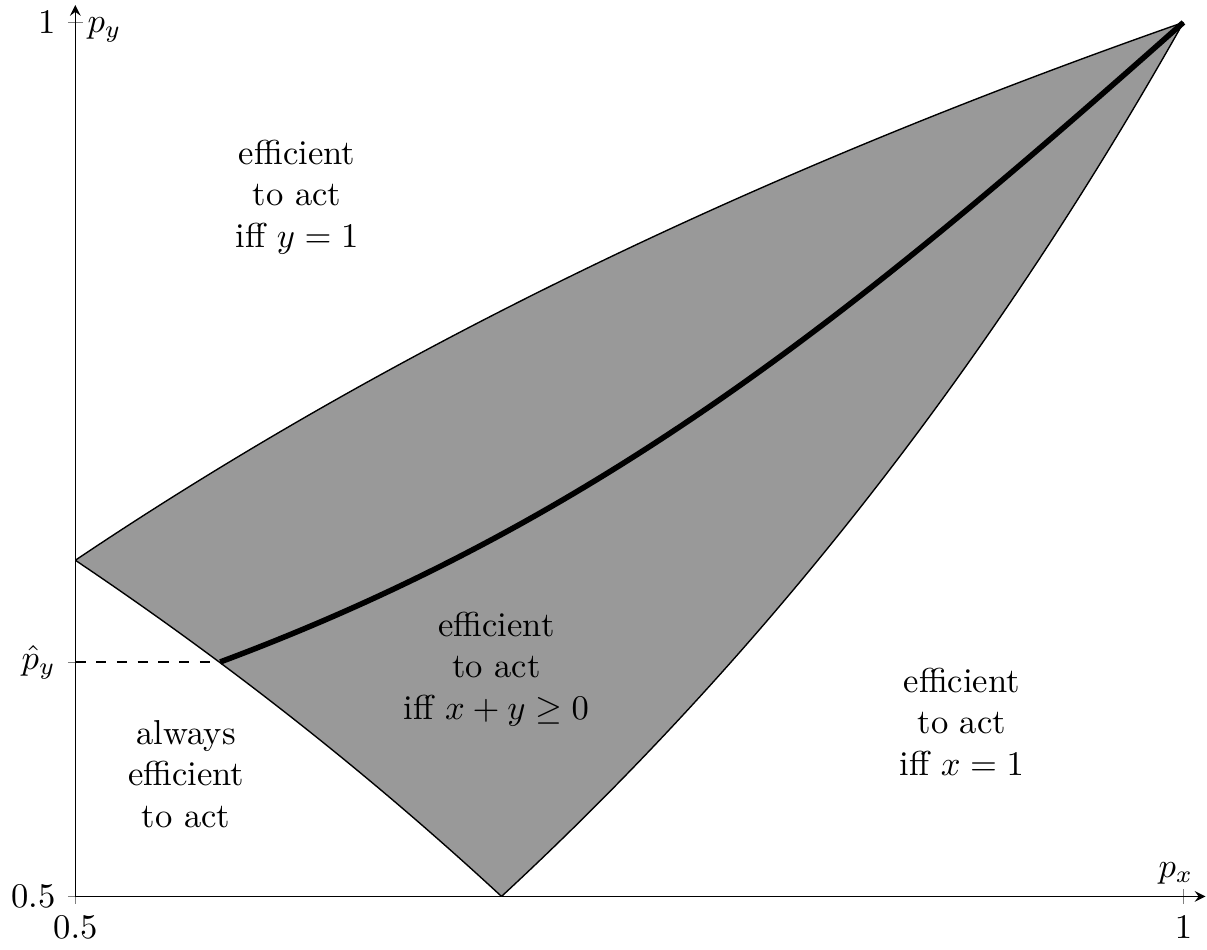}
\caption{\emph{Critical values of the quality of information.} The shaded area is the parameter region ($p_x,p_y$) in which it is efficient to act iff $x+y\geq 0$ (our baseline case). On the top left of the shaded region, it is efficient to act iff $y\geq 0$; and on the bottom right iff $x \geq 0$. The bottom left is the area in which even two negative signals cannot overturn the prior $\beta$ and it is always efficient to act. The thick black line depicts the beliefs at which $F^b=F^u$ and  ($p_x^\ast,p_y^\ast$). Changes in $p_x$ represent movements parallel to the $x-$axis; changes in $p_y$ represent movements parallel to the $y-$axis. Welfare drops for horizontal moves crossing the black line (see Figure \ref{fig:welfare}). In this example, $\beta=9/13$.}\label{fig:regions}
\end{figure}

\subsection{Signal Precision} 
\label{sub:comparative_statics}
Suppose that the verifiable information becomes more precise; that is, $p_x$ increases to some $p_x'>p_x$. By Lemma \ref{lem:FdminusFhincreasing}, we could have $F^b < F^u$ at $p_x$ and $F^b > F^u$ at $p_x'$. Define the following \emph{critical threshold} of information quality.

\begin{definition}
The precision level $p_x^\ast$ is a critical threshold of information quality given $(p_y, \beta)$ if the following two conditions hold:
\begin{enumerate}
\item The critical punishment levels are equal: $F^b(p_x^* , p_y,\beta) = F^u(p_x^*,p_y,\beta)$ (see \eqref{Equation:Fd and Fh}). 
\item The informational environment $S=(p_x^*,p_y,\beta)$ is in the interior of environments in which it is efficient to act if and only if $x+y \ge 0$. 
\end{enumerate} 
\end{definition}

Figure \ref{fig:regions} sketches these levels for a fixed $\beta$ in the $(p_x,p_y)$ plane. It is efficient to act if and only if $x+y\geq 0$ inside the shaded region. The thick black line plots the critical information quality,  $p_x^*(p_y)$.

With some abuse of notation, let $W^*(p_x)$ [resp. $W^*(p_y)$] denote the (ex ante) welfare corresponding to the welfare-maximizing equilibrium for some precision level $p_x$ [resp. $p_y$] in an otherwise-fixed environment $(p_y,\beta,\gamma)$ [resp. $(p_x,\beta,\g)$].

\begin{proposition}\label{prop:main_result}An increase in the precision of the verifiable signal can reduce the welfare in non-knife-edge cases. Formally, if $p_x^*$ is a critical threshold, then there is an $\epsilon>0$ such that $W^*(p_x) > W^\ast(p_x')$ whenever $p_x^* - \epsilon < p_x < p_x^* < p_x' <  p_x^*+\epsilon$.\end{proposition}

Proposition \ref{prop:main_result} is driven by the sign change of $F^b-F^u$ around $p_x^\ast$ as described in Lemma \ref{lem:FdminusFhincreasing}. For example, suppose that $\g$, the prior probability of the agent being unbiased, is high and $p_x$ is slightly below $p_x^*$. Here, since $F^b < F^u$, the optimal equilibrium is as in Table \ref{Table:Fh less than Fd and Fbar =Fd}. In particular, it is possible to have the biased agent act with probability less than one on $(-1,-1)$ while having the unbiased agent act with probability one on $(-1,1)$. 

Increasing $p_x$ to slightly above $p_x^*$ implies that $F^b>F^u$. We can no longer have the biased agent act on $(-1,-1)$ with probability less than one while having the unbiased agent act with a positive probability on $(-1,1)$. Therefore, the designer is left with two options. Either she gives a universal free pass, or she achieves partial deterrence of the biased agent at the cost of a chilling effect on the unbiased agent.

While the above discussion focuses on the negative effect of improving the verifiable information, there is also a positive effect. 
An increase in $p_x$ implies that, conditional on $\theta=1$, realization $x=1$ occurs more often\textemdash an improvement in welfare. Yet the effect of such an improvement is continuous in $p_x$, while the effect due to a regime change, from $F^b < F^u$ to $F^b > F^u$, is discrete. Therefore, welfare declines discretely.

\begin{figure}[h!]

  \centering
  \begin{subfigure}{.45\textwidth}
\includegraphics[width=\textwidth]{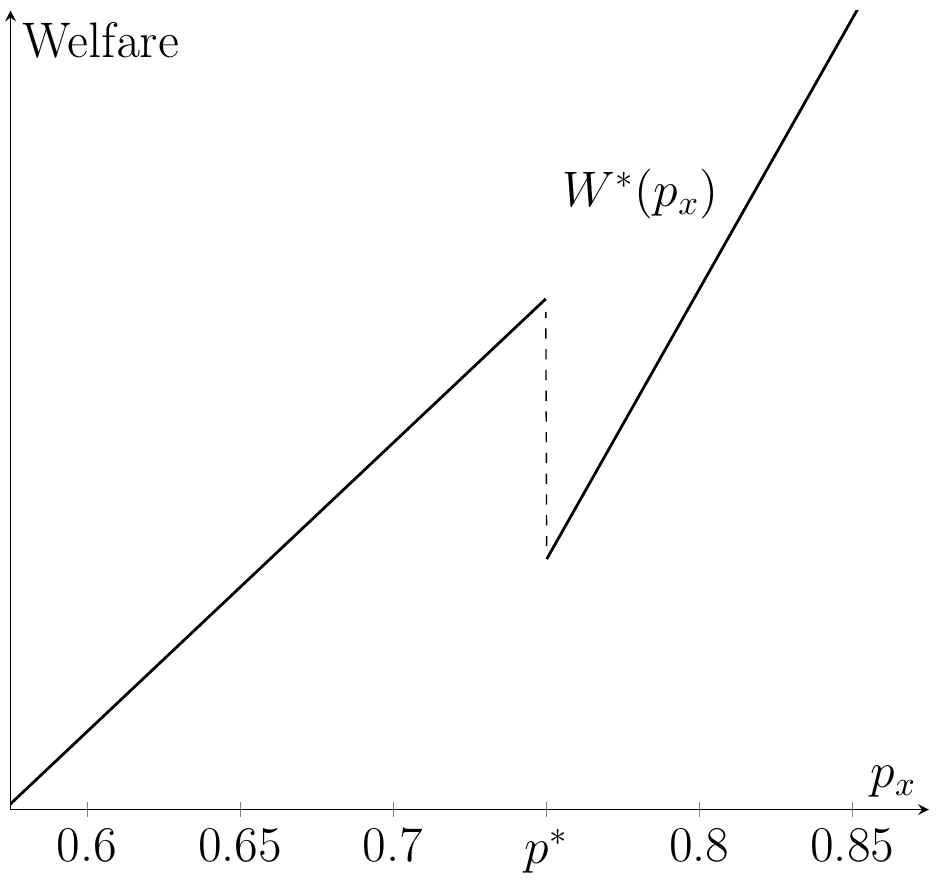}
\caption{Welfare in precision of $\mathbf{X}$}
   \end{subfigure} \hfill
  \begin{subfigure}{.45\textwidth}
  \includegraphics[width=\textwidth]{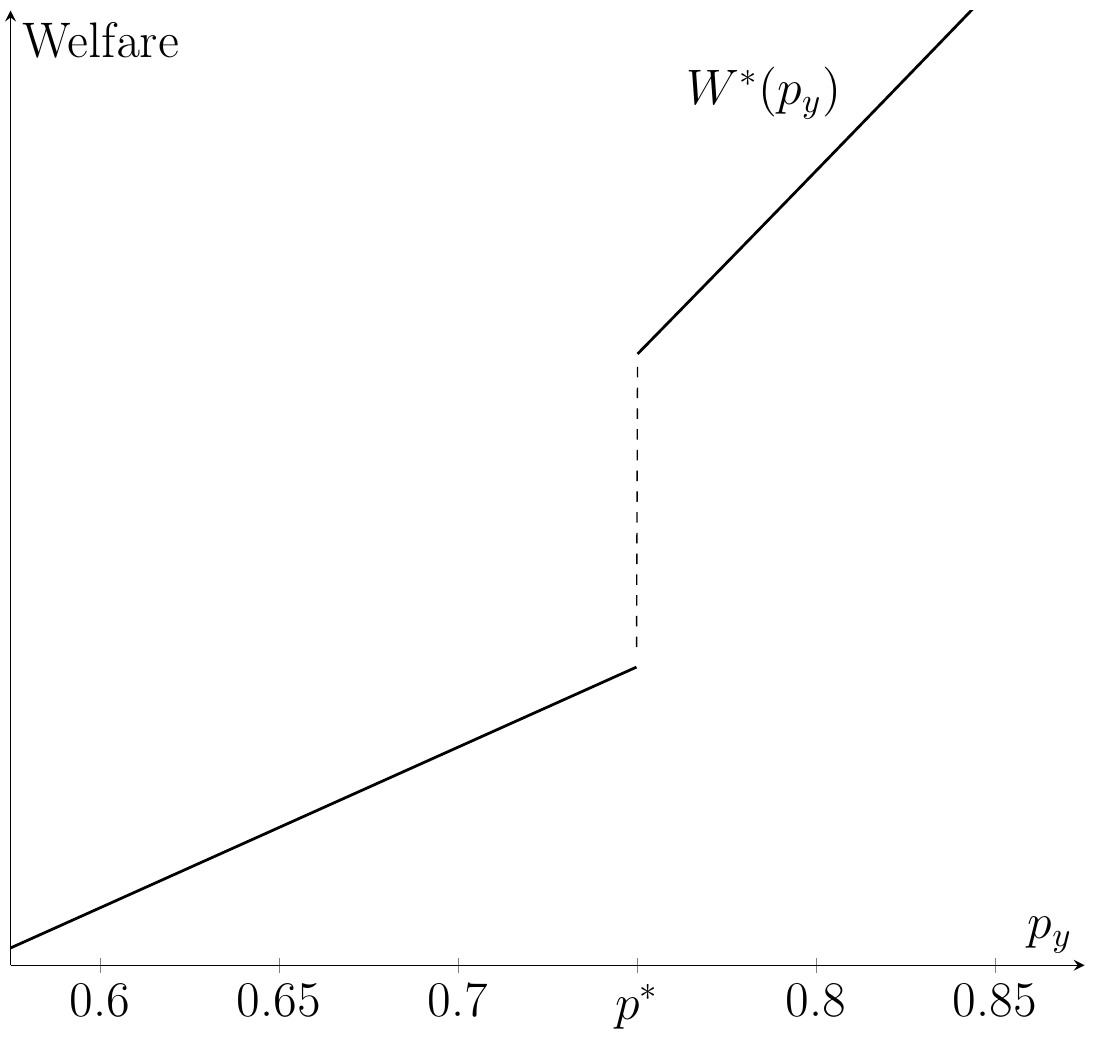} 
\caption{Welfare in precision of $\mathbf{Y}$}
  \end{subfigure}
\caption{\emph{Welfare for different precision levels.} Welfare is depicted as a function of the precision of the verifiable information, $W^\ast(p_x)$ (left panel), and as a function of the precision of the unverifiable information, $W^\ast(p_y)$ (right panel). The discontinuity is at the point at which $F^u=F^b$ such that we switch from the bottom row to the top row of Table \ref{Table: equilibrium strategy profiles} (left panel) or from the top row to the bottom row (right panel). In the entire domain of information qualities pictured, acting is efficient iff $x+y\geq 0$. Also, the maximum punishment, $\overline{F}$, is chosen optimally throughout. Parameters: $\overline{\gamma}=1/2,\gamma=11/20,\beta=9/13$, and $p_y=3/4$ (left panel), $p_x=3/4$ (right panel).}\label{fig:welfare}
\end{figure}

We want to emphasize that Proposition \ref{prop:main_result} gives a local comparative static. A sufficiently large increase of $p_x$ increases welfare. For example, for a fixed $p_y$,  as $p_x \rightarrow 1$, heavily punishing the agent for any failure implies that the project is implemented if, and only if, it is good.
In panel (a) of Figure \ref{fig:welfare}, we display welfare as a function of the precision of the verifiable information, $p_x$. Precisely at the critical threshold of information quality, $p_x^*$, we see a discontinuous decrease in it as a result of changes in the optimal punishment. For verifiable signals less informative than $p_x^*$, the optimal punishment can partially deter the biased agent without inducing any chilling effect. 
In contrast, for verifiable signals more informative than $p_x^*$, to deter the biased agent implies a complete chilling effect. 

It is tempting to think that the same comparative static holds for the precision of the unverifiable signal. This naive reasoning turns out to be false. 
\begin{samepage}
\begin{proposition}\label{prop:pyincreasegood}
An increase in the precision of the unverifiable signal always increases welfare. That is, $W^\ast(p_y')\geq W^\ast(p_y)~\forall p_y'>p_y .$
\end{proposition}\end{samepage}

The main difference between $p_x$ and $p_y$, and the driver of our results, lies in their effect on 
$F^b - F^u$ as seen in Lemma \ref{lem:FdminusFhincreasing}. The main conflict in our environment is that, at times, we want the unbiased agent to decide in favor of undertaking the project despite negative verifiable information, $x=-1$. We want him to rely on the positive unverifiable information he received. However, the associated cost is that---because of the lack of punishment---the biased agent undertakes a project even if $x=y=-1$ (that is, all the information is against it). Increasing the precision of the unverifiable information helps the designer. With increased $p_y$, $y=1$ suggests a higher likelihood of the project quality being good. Therefore, it makes the unbiased agent more confident about undertaking the project when $x=-1$ but $y=1$. At the same time, it disincentivizes the biased agent from undertaking the project when $x=y=-1$. Thus, welfare increases.

In contrast, increasing the precision of verifiable information disincentivizes both types given negative verifiable information. It increases the threat of punishment, as it indicates a higher chance of failure and thus of punishment if the project is undertaken. In addition, and only for the unbiased type, there is a second deterring force. His payoff is connected to the success of the project directly, and the incentives to undertake the project, given $x=-1$, decline in the precision of $\pubsigr$, regardless of the punishment. Because of these two effects, increasing the precision of the verifiable signal may lead to lower welfare.

In Panel (b) of Figure \ref{fig:welfare}, we display welfare as a function of the precision of the unverifiable information, $p_y$. As in panel (a), at the critical threshold of information quality, $p_y^*$, there is a jump in welfare. However, in contrast to Panel (a), the jump is upward. As the precision of the  unverifiable information exceeds the critical threshold, we move from a situation of full deterrence paired with a full chilling effect to the better situation of partial deterrence with no chilling effect.  

\section{Extensions}\label{sec:Robustness}
Given the simplicity of our mechanism, it is natural to wonder about the generality of the forces that lead to the different welfare implications of improving the verifiable and unverifiable information. Do these forces hinge on the intricate details of a formal legal system? Do they rely on the specific assumptions we made about the signal structure? 

In this section, we address these questions by highlighting the robustness of our main message to different model specifications. We begin with an abstract principal-agent model. Then we change the objective of the court: what if it aimed to punish the agent for acting against his better knowledge (objective means rea)? Next, we consider punishment for inaction: what if the court can punish the agent for not acting? We also extend the model to more than two types of agents, and we discuss a richer signal structure.

\subsection{A Contracting Model} 
\label{sub:exogenous_maximum_punishment}

We temporarily leave the legal setting with its three players (the designer, the agent, and the court) and consider an abstract model with a principal and an agent, essentially combining the designer and court into a single principal. We consider two versions. First, the principal commits to a punishment rule \emph{ex ante} (the commitment case). That is, the principal designs and commits to a punishment scheme  before the agent has acted. Second, the principal decides \emph{ex post} and without constraints on the punishment---that is, after the agent has acted (the ex post screening case). We show that our results continue to hold in both  cases. 

\paragraph{Commitment.} There is a principal and an agent. Nature moves first and draws $\theta,\omega,x,y$ according to a commonly known informational environment, $(S,\gamma)$. The principal observes $x$. Thereafter she commits to a punishment $F: \supp(\pubsigr) \rightarrow \mathbb{R}_+$. The agent observes $F$, his type $\omega$, and $(x,y)$. The agent selects $a \in \{0,1\}$. If $a \theta=-1$, the agent gets (in addition to his gross payoff $u^\omega(a,\theta)$) punished by $F(x)$.

Given $F$, the problem of the agent is identical to that in the baseline case. Therefore, $F^b$ and $F^u$ are the same as in the baseline model, which, in turn, implies that Lemma \ref{lem:FdminusFhincreasing} holds. 
Moreover, it remains without loss to consider as a candidate for an optimal fine $F \in\{F^b,0\}$ when $F^b>F^u$ and $F\in\{F^b,F^u\}$ when $F^b<F^u$.

The only departure from the baseline model is that the principal need not be indifferent in punishing the agents. The version of Table \ref{Table: equilibrium strategy profiles} from the baseline, adapted to the commitment case, is depicted in Table \ref{Table: Committment}.

\begin{table}[tb]
\caption{Strategy profiles in the optimal equilibria}
\label{Table: Committment}
\begin{subtable}[t]{0.95\linewidth}
    \caption*{When $F^b > F^u$}
    \label{Table (contracting): Fd greater than Fh}
    \centering
\begin{subtable}[t]{0.45\linewidth}
      \caption{When $F(-1) = 0$}
              \label{Table: (contracting) Fh less than Fd and Fbar =0}
      \centering
         \begin{tabular}{lll}
           $(x,y)$ &  $a^u$ & $a^b$\\
           \hline
           (-1,-1) & 0 & 1 \\
           (-1,1) & 1 & 1
        \end{tabular}
\end{subtable}
\begin{subtable}[t]{0.45\linewidth}
      \caption{When $F(-1) = F^b$}
              \label{Table: (contracting) Fh less than Fd and Fbar =Fd}
      \centering
         \begin{tabular}{lll}
           $(x,y)$ &  $a^u$ & $a^b$\\
           \hline
           (-1,-1) & 0 & 0 \\
           (-1,1) & 0 & 1
        \end{tabular}
\end{subtable}
\end{subtable}
\bigskip
\begin{subtable}[t]{0.95\linewidth}
    \caption*{When $F^u > F^b$}
    \label{Table (contracting): Fd less than Fh}
    \centering
\begin{subtable}[t]{0.45\linewidth}
       \caption{When $F(-1) = F^b$}
              \label{Table: (contracting) Fh greater than Fd and Fbar =Fd}
      \centering
         \begin{tabular}{lll}
           $(x,y)$ &  $a^u$ & $a^b$\\
           \hline
           (-1,-1) & 0 & 0 \\
           (-1,1) & 1 & 1
        \end{tabular}
\end{subtable}
\begin{subtable}[t]{0.45\linewidth}
     \caption{When $F(-1) = F^u$}
              \label{Table: (contracting) Fh greater than Fd and Fbar =Fh}
      \centering
         \begin{tabular}{lll}
           $(x,y)$ &  $a^u$ & $a^b$\\
           \hline
           (-1,-1) & 0 & 0 \\
           (-1,1) & 1 & 1
        \end{tabular}
\end{subtable}
\end{subtable}
\end{table}

On the one hand, if $F^u>F^b$, committing to $F(-1)=F^b$ guarantees that the agent takes the (interim) efficient action regardless of his type. On the other hand, if $F^b>F^u$, the payoffs are identical to those in the baseline case. The unbiased agent is too pessimistic about the state and thus is completely chilled by any fine that deters the biased agent. The principal has to decide whether she prefers to prevent the biased agent at the cost of chilling the unbiased agent or to avoid the chilling effect at the expense of no deterrence. Regardless, we lose interim efficiency. 

As the environment changes, from $F^b<F^u$ to $F^b>F^u$, welfare suffers a discrete loss because of the inability to implement the interim efficient action, which outweighs the marginal gain from better information. Because Lemma \ref{lem:FdminusFhincreasing} applies, we move---in Table \ref{Table: Committment}---from the bottom row to the top row at $p_x^\ast$ as $p_x$ increases and from the top row to the bottom row at $p_y^\ast$ as $p_y$ increases. 

In summary, in the commitment case, welfare always improves in the precision of the unverifiable information, while it may decline in the precision of the verifiable information, just like in our main results. Moreover, the driving intuition remains the same in this case.

\paragraph{Ex Post Screening.} There are a principal and an agent. Nature moves first and draws $\theta,\omega,x,y$ according to the commonly known informational environment $S$. Then the agent observes $\omega, x,y$. Thereafter, the agent selects $a \in \{0,1\}$. If $a \theta=-1$, the principal observes $x$ and can inflict a punishment $F \in \mathbb{R}_+$ on the agent that reduces his gross payoff from acting, $u^\omega(\cdot)$, by $F$. The principal receives a benefit of $F$ if she punishes a biased agent and suffers a loss $LF$ if she punishes an unbiased agent.

As in the baseline setting, the principal's preferences determine a threshold $\overline{\gamma}$ such that the principal wants to punish if her belief $\gamma_{x}$, conditional on $a\theta=-1$ and realization $\pubsigr=x$, is less than $\overline{\gamma}$. Similarly, she wants to acquit if $\gamma_{x}>\overline{\gamma}$.

We make two observations. First, $\gamma_{x}<\overline{\gamma}$ cannot be an on-path belief. If it were, the principal would select $F(x)=\infty$, which, in turn, would lead to full deterrence.  
Second, if a free pass is not universally optimal, it cannot be an equilibrium outcome. If it were, the principal's belief would be $\gamma_{-1}<\overline{\gamma}$, which implies punishment---a contradiction.

The two observations imply that the principal either implements full deterrence or has to be indifferent in any equilibrium. If $F^b>F^u$, full deterrence is the only option, whereas when $F^u>F^b$, full deterrence cannot be optimal. Consequently, the equivalent to Table \ref{Table: equilibrium strategy profiles} for this case is Table \ref{Table exogenous: equilibrium strategy profiles}.

\begin{table}[tb]
\caption{Strategy profiles in the optimal equilibria}
\label{Table exogenous: equilibrium strategy profiles}
\begin{subtable}[t]{0.95\linewidth}
    \caption*{When $F^b > F^u$}
    \label{Table exogenous: Fd greater than Fh}
    \centering
\begin{subtable}[t]{0.45\linewidth}
      \caption{When $\E[F] \geq F^b$}
              \label{Table exogenous :Fh less than Fd and Fbar =0}
      \centering
         \begin{tabular}{lll}
           $(x,y)$ &  $a^u$ & $a^b$\\
           \hline
           (-1,-1) & 0 & 0 \\
           (-1,1) & 0 & 0
        \end{tabular}
\end{subtable}
\end{subtable}
\bigskip
\begin{subtable}[t]{0.95\linewidth}
    \caption*{When $F^u > F^b$}
    \label{Table exogenous: Fd less than Fh}
    \centering
\begin{subtable}[t]{0.45\linewidth}
       \caption{$\E[F] = F^b$}
              \label{Table exogenous:Fh greater than Fd and Fbar =Fd}
      \centering
         \begin{tabular}{lll}
           $(x,y)$ &  $a^u$ & $a^b$\\
           \hline
           (-1,-1) & 0 & $\eta^b$ \\
           (-1,1) & 1 & 1
        \end{tabular}
\end{subtable}
\begin{subtable}[t]{0.45\linewidth}
     \caption{$\E[F] = F^u$}
              \label{Table exogenous:Fh greater than Fd and Fbar =Fh}
      \centering
         \begin{tabular}{lll}
           $(x,y)$ &  $a^u$ & $a^b$\\
           \hline
           (-1,-1) & 0 & 0 \\
           (-1,1) & $\eta^u$ & 1
        \end{tabular}
\end{subtable}
\end{subtable}
\end{table} 
In Table \ref{Table exogenous: equilibrium strategy profiles}, $\eta^b$ and $\eta^u$ are such that the principal is indifferent. Being indifferent, the principal can select any punishment scheme. However, to make the agent indifferent as well, we need it to be true that the expected punishment $\E[F]=F^b$ or $\E[F]=F^u$.

The ex post screening case strengthens our results. In Table \ref{Table exogenous: equilibrium strategy profiles}, welfare is strictly lower in the top row compared to the baseline. The reason is the following: In this environment, we lack a designer to optimally limit the punishment ex ante. 
The bottom row, however, yields the same welfare as in the baseline case. We see that the designer's ability to limit the punishment is beneficial, particularly when $F^b > F^u$, which occurs when the verifiable signal is precise. 

\paragraph{Relationship to the Baseline.} The commitment case corresponds to strict liability in the legal setting. In certain situations---for example, if the realization of the verifiable information is some specific $x$---an action makes the agent liable ``per se.'' That is, the court does not form an opinion about the agent's type but punishes based only on $a,\theta,x$. Such a case is directly captured by the commitment model.

The ex post screening case encompasses scenarios in which the magnitude of the punishment is exogenous; for example, the agent gets fired from his job. In some of the examples we discuss in Section \ref{sub:examples}, such an exogenous punishment appears appropriate.

\subsection{Objective Mens Rea}\label{Subsection: mens rea different interpretation}

In this section, we address the question of how relevant the assumption of  subjective mens rea is to our substantive results. To that end, we  present an extension in which the court follows objective mens rea instead.

Our running assumption in the baseline model is that the court's objective is to infer the \emph{agent's preferences} from the information available to it and it wants to punish only the biased agent. 
That is, it wants to punish an agent only if it is sufficiently convinced that the agent caused harm because his preferences are not aligned with society's. Legal philosophers call this notion subjective mens rea.

An alternative specification could be to assume that the court tries to infer the agent's \emph{(nonverifiable) information} from what it observes: the choice made by the agent, the outcome, and the (verifiable) information. And the court wants to punish the agent for acting when the available information indicated that he should have exercised restraint. Legal philosophers call this notion objective mens rea.

While mens rea as a requirement for conviction is a doctrine from  criminal law, it serves as a principle in tort cases too. That is, a person's type or intentions play an important role in courts' conviction decisions in tort cases as well. For example, standards of care such as recklessness and (gross) negligence focus on the conscious and voluntary state of mind. In particular, if the court employs the reasonable-person standard to assess the presence of negligence, then its goal is to determine whether a person with reasonable preferences would have acted in a certain way.\footnote{The reasonable-person standard explicitly takes into account that the reasonable person is sophisticated and acts ``in the shadow of the law.'' That is, she takes legal consequences into account when deciding whether to act.}

In addition, discrimination lawsuits also use type attributes to prove intentional discrimination under Title VII of the Civil Rights Act. For example, in Wilson v. Susquehanna Township Police Department, 55 F.3rd 126 (3rd Cir. 1995),\footnote{See \url{https://m.openjurist.org/55/f3d/126/wilson-v-susquehanna-township-police-department-l}.}  the court ruled that the police chief's intent was to discriminate because it was evident (to the court) that the chief held a ``strong gender bias.'' The court did not question the lower court's ruling that there may have been reasons to promote another person instead of the plaintiff but overruled it on the basis of the ``discriminatory attitude'' of the chief as ```direct evidence' of discriminatory animus.''\footnote{In some cases, the court even uses prior acts to determine the agent's type; see, for example, \url{https://www.newyorker.com/magazine/2012/03/19/tax-me-if-you-can}, about a case in which a citizen was acquitted because of prior proof of character. For an economic discussion on the use of character evidence in various settings, see \citet{lester2012information,BullWatson219,10.2307/1123746}. In our model, character evidence (as usually defined) is absent. Any information the court uses to determine culpability is either  about the \emph{project} or about the agent's equilibrium behavior.}

As discussed above, both subjective and objective mens rea seem to be  reasonable assumptions depending on which environment is being captured. We choose to use subjective mens rea in the baseline model for two reasons. The first is an economic reason: we are interested in the welfare-maximizing equilibria. As we show later,  welfare under subjective mens rea is greater than under objective mens rea. The second reason is that, as discussed above, the courts seem to employ subjective mens rea regularly in and outside of criminal law. Having said that, we want to highlight that our main comparative statics (and the underlying intuition) hold regardless of which formulation of mens rea is used. We show this below. 

\paragraph{Objective mens rea.}
Let the agent be punished if he took an action, $a=1$, that resulted in a bad outcome, $\theta=-1$, and the court is sufficiently convinced that the agent's signal indicated that he should not have acted; that is, the agent's signal was $(-1,-1)$. 
That is, under objective mens rea the court punishes if \[q := \prob(\pvtsigr=1 \vert \theta = -1, a= 1, \pubsigr=-1) \le \overline{\gamma}.\]  

Fixing all the parameters we obtain our first result.

\begin{proposition}\label{prop:welfarelarger}

Expected welfare in the optimal equilibrium is weakly higher if the court employs subjective mens rea than if it employs objective mens rea.
\end{proposition}

The intuition underlying Proposition \ref{prop:welfarelarger} is seen from 
Table \ref{Table extension: equilibrium strategy profiles}. There are three main differences in the equilibrium behavior compared to the baseline case: (i) if $F^b>F^u$ and $F=F^b$, the biased agent acts with positive probability $\eta_1$ (as opposed to zero probability in the baseline case) on $(-1,-1)$; (ii) if $F^b<F^u$, the optimal punishment is always $F^b$; and (iii) the probability with which the biased agent acts on $(-1,-1)$ 
is $\eta_2$, which is larger than $\eta^b$, used in the baseline case. These three properties 
imply Proposition \ref{prop:welfarelarger}.

\begin{table}[tb]
\caption{Strategy profiles in the optimal equilibria}
\label{Table extension: equilibrium strategy profiles}
\begin{subtable}[t]{0.95\linewidth}
    \caption*{When $F^b > F^u$}
    \label{Table: obj MR Fd greater than Fh}
    \centering
\begin{subtable}[t]{0.45\linewidth}
      \caption{When $\of = 0$}
              \label{Table extension:Fh less than Fd and Fbar =0}
      \centering
         \begin{tabular}{lll}
           $(x,y)$ &  $a^u$ & $a^b$\\
           \hline
           (-1,-1) & 0 & 1 \\
           (-1,1) & 1 & 1
        \end{tabular}
\end{subtable}
\begin{subtable}[t]{0.45\linewidth}
      \caption{When $\of = F^b$}
              \label{Table extension:Fh less than Fd and Fbar =Fd}
      \centering
         \begin{tabular}{lll}
           $(x,y)$ &  $a^u$ & $a^b$\\
           \hline
           (-1,-1) & 0 & $\eta_1$ \\
           (-1,1) & 0 & 1
        \end{tabular}
\end{subtable}
\end{subtable}
\bigskip
\begin{subtable}[t]{0.95\linewidth}
    \caption*{When $F^u > F^b$}
    \label{Table extension: Fd less than Fh}
    \centering
\begin{subtable}[t]{0.45\linewidth}
       \caption{When $\of = F^b$}
              \label{Table extension:Fh greater than Fd and Fbar =Fd}
      \centering
         \begin{tabular}{lll}
           $(x,y)$ &  $a^u$ & $a^b$\\
           \hline
           (-1,-1) & 0 & $\eta_2$ \\
           (-1,1) & 1 & 1
        \end{tabular}
\end{subtable}
\end{subtable}
\end{table}

\begin{figure}[t]
  \centering
  \begin{subfigure}{.45\textwidth}
 \includegraphics[width=\textwidth]{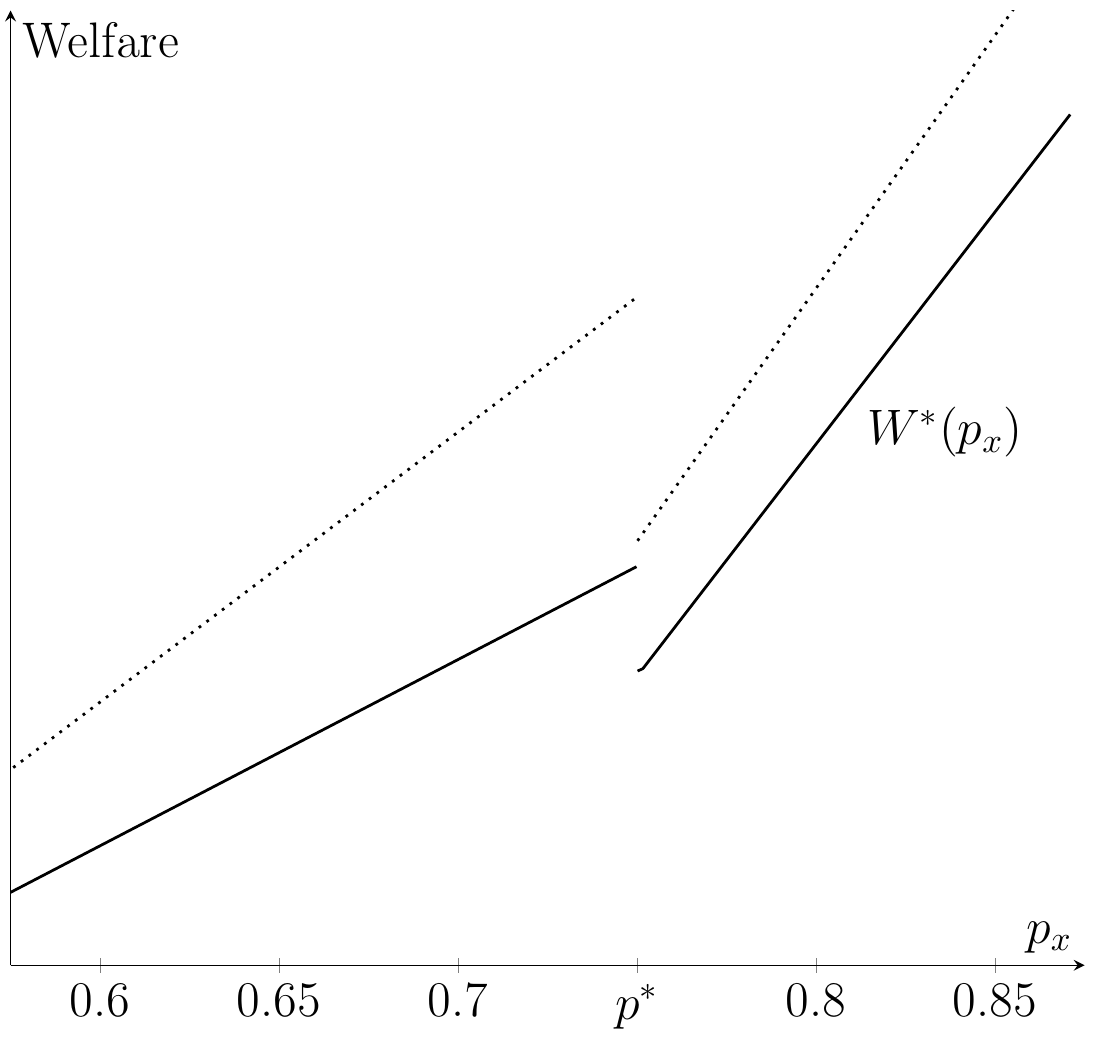}
\caption{Welfare changes in precision of $\mathbf X$}
   \end{subfigure} \hfill
  \begin{subfigure}{.45\textwidth}
  \includegraphics[width=\textwidth]{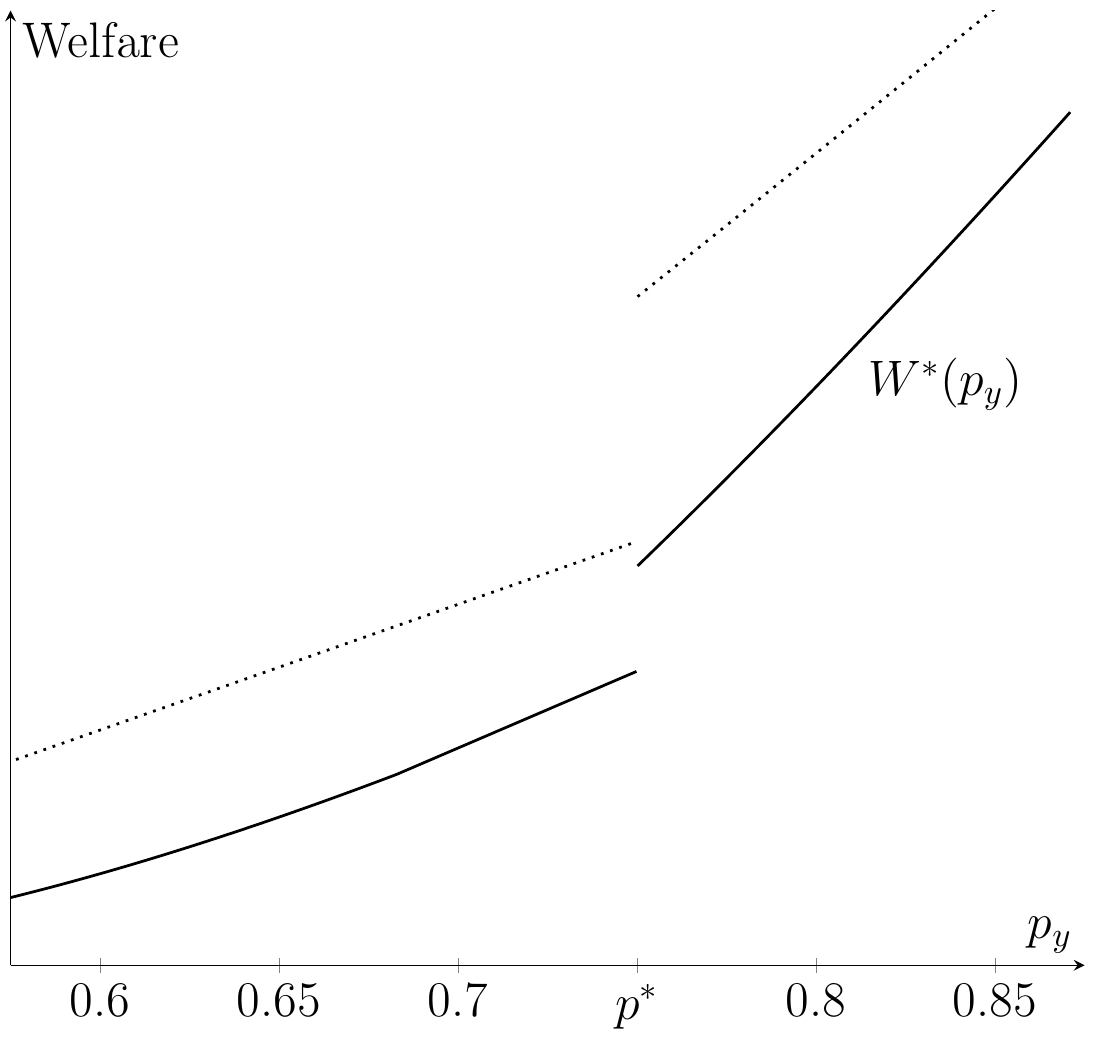} 
\caption{Welfare changes in precision of $\mathbf Y$}
  \end{subfigure}
\caption{\emph{Welfare when the court aims to convict only agents that acted despite better information.} Welfare is depicted as a function of the precision of the verifiable information, $W^\ast(p_x)$ (left panel), and as a function of the precision of the unverifiable information, $W^\ast(p_y)$ (right panel). Solid lines depict the values under objective mens rea. Dotted lines are the values for the baseline case of subjective mens rea. Parameters: $\overline{\gamma}=1/2,\gamma=11/20,\beta=9/13$, and $p_y=3/4$ (left panel), $p_x=3/4$ (right panel).}\label{fig:welfareD}
\end{figure}

We now present the effects of changes in information quality for the alternative specification of the court's objective. 
Here, unlike in the baseline model, welfare may decline upon improving the precision of $\pvtsigr$, the unverifiable information, when the court adopts objective mens rea. If such a decline occurs, it occurs at the critical level $p_y^\ast$. Table \ref{Table extension: equilibrium strategy profiles} highlights the underlying reason. As $p_y$ increases, we may move from panel (b) to panel (c). That transition implies less deterrence of the biased agent, $\eta_2>\eta_1$, but removes the chilling effect on the unbiased agent. Thus, welfare declines in the transition only if the former effect is larger than the latter.

Other than at the critical threshold $p_y^\ast$, welfare is increasing in $p_y$. The following condition provides a necessary and sufficient condition to guarantee that the positive effect of increasing $p_y$ that mitigates the chilling effect outweighs the negative effect of reduced deterrence: 

\begin{align}
\frac{1-\gamma}{\gamma} (\eta_2(p_y^\ast)-\eta_1(p_y^\ast)) \le \frac{\beta p_y (1-p_x) - (1-\beta) p_x(1-p_y)}{(1-\beta)p_x p_y - \beta (1-p_x)(1-p_y)}\label{Equation: Condition from objective mens rea for increasing welfare in y}
\end{align}

The following proposition provides sufficient conditions such that Propositions \ref{prop:main_result} and \ref{prop:pyincreasegood} maintain for objective mens rea. To state it, we need to introduce some additional notation. Let $\mc Y(\beta,p_x)$ be the set of $p_y$'s such that it is efficient to act iff $\max\{x,y\}=1$ when the precision of $\pubsigr$ is $p_x$ and that of $\pvtsigr$ is $p_y$. Notice that $\mc Y(\beta,p_x)$ is compact, and hence we can define $\underline{p_y}(\beta,p_x) := \min \mc Y(\beta,p_x)$ and $\overline{p_y}(\beta,p_x) := \max \mc Y(\beta,p_x)$. 
\begin{proposition}\label{result_objectivemensrea}
Suppose that it is efficient to act if and only if at least one signal is positive, $\max\{x,y\}=1$, and that the court employs objective mens rea. 
\begin{enumerate}
\item \textbf{Precision of }$\boldsymbol X$: Consider an increase in the precision from $p_x$ to $p_x'>p_x$. Proposition \ref{prop:main_result} applies. That is, welfare at $p_x'$ may be lower than at $p_x$. 
\item \textbf{Precision of }$\boldsymbol Y$: Consider an increase in the precision from $p_y \in \mathcal Y(\beta,p_x)$ to $p_y'\in \mathcal Y(\beta,p_x)>p_y$. Proposition \ref{prop:pyincreasegood} applies---welfare is unambiguously higher at $p_y'$ than at $p_y$---if either of the following is true: \begin{itemize}
    \item Condition \eqref{Equation: Condition from objective mens rea for increasing welfare in y} holds.
    \item $p_y^\ast \notin \mathcal Y(\beta,p_x)$. 
\end{itemize} 
\end{enumerate}
 \end{proposition}

There are two economically interpretable sufficient conditions that arise from \eqref{Equation: Condition from objective mens rea for increasing welfare in y}. First, if $\g$---the proportion of  unbiased agents in the society---is sufficiently large, then the society prefers not to deter the relatively few biased types from acting on $(-1,-1)$. The reason is the associated cost of chilling  the unbiased types. So when $F^b > F^u$, the society prefers to give a free pass. However, as $p_y$ increases, we have $F^b < F^u$, leading to an increase in welfare, as in Proposition \ref{prop:pyincreasegood}. 

Another sufficient condition relates to the tolerance of the court. If $1- \overline{\gamma} > p_y^*$---that is, the court convicts when it is sufficiently confident that the agent acted on all negative information---then an increase in $p_y$ leads to an easier separation of the two types for the court. The reason is that the information effect is similar to the one in the baseline model. These and other conditions, and the reasoning behind them, are detailed in Appendix \ref{Appendix: Objective mens rea}.

\subsection{Punishment for Inaction}\label{Section: conviction upon a=0}
Throughout the paper, we have assumed that the court can punish the agent only when $a=1$ and $\theta=-1$. Our choice is motivated mainly by realism \citep[for recent experimental evidence, see][]{cox2017status}.\footnote{ Scholars debate  whether not punishing  inaction stems from a cognitive bias or from rational behavior \citep[for an overview, see][]{woollard2019doing}.} 

We now extend our model to allow the court to punish the agent for not acting. 
That is, suppose that the court always sees $\theta$ and $x$ regardless of whether  the agent acted. The court would ideally like to punish the agent for displaying excessive caution by not acting. However, given the lack of commitment on our court's part, this ability to punish for inaction does not remedy the issue. To see this, first recall that, regardless of the punishment scheme, for any realization of the unverifiable signal that an unbiased agent acts on with strictly positive probability, a biased agent finds it optimal to act. Therefore, for any realization, inaction only increases the likelihood that the agent is unbiased, and the court's posterior over the agent's being unbiased must be weakly higher than its prior. Because the prior is higher than the conviction cutoff---that is, $\g>\overline{\gamma}$---by assumption, the court chooses to not punish the agent for inaction, even if allowed to do so. 

\subsection{More than Two Types of Agents} 
\label{sub:more_than_two_types_of_agents}

We now  extend our model to capture a setting in which the agent's preferences can have various degrees of misalignment with the designer's preferences. Specifically, suppose that there is a finite set of types, $\{1,2,\ldots, K\}$. The utility of a type $k$ agent is given by $u^k(a,\theta) := a[\lambda^k \theta + (1-\lambda^k)]$, where $\lambda^k \in [0,1]$. Suppose that $0= \lambda^1 \le \lambda^2 \ldots \lambda^K = 1$. Notice that type $1$ is the biased agent in our main model, while type $K$ is the unbiased agent. Let $\mu_k$ denote the ex ante probability that the agent is of type $k$. Even in this environment, the essential problem we face is the same: we want to define the maximum punishment that incentivizes agents to act on $(-1,1)$ while disincentivizing  agents from acting on $(-1,-1)$. Given any $F$, the following fact is immediate: 
$$a^k(-1,\cdot) > 0 \implies a^m(-1,\cdot) =1\quad \forall m < k.$$ 
Therefore, we can define $K^1$ to be the highest type that acts on $(-1,1)$ and $K^{-1}$ to be the highest type that acts on $(-1,-1)$. Notice that $\lm^{K^1} \ge \lm^{K^{-1}}$. 

This model delivers the same results as Propositions \ref{prop:main_result} and \ref{prop:pyincreasegood}. To see why, recall that the main driver of those results is Lemma \ref{lem:FdminusFhincreasing}, which established that $F^b- F^u$ is increasing in $p_x$ and decreasing in $p_y$. Similarly to $F^b$ and $F^u$, we can define $F_1^k$ to be the largest fine that allows type $k$ to act on $(-1,1)$ and $F_{-1}^k$ to be the minimum fine required to deter type $k$ from acting on $(-1,-1)$. Then, straightforward algebra (analogous to the expression of $F^b-F^u$) yields 
\begin{align*}
F^{K^{-1}}_{-1} - F^{K^1}_1 = -2(\lm^{K^1}-\lm^{K^{-1}}) + \frac{\beta}{1-\beta}\frac{1-p_x}{p_x}\left[\frac{1-p_y}{p_y} - \frac{p_y}{1-p_y}\right].
\end{align*}
Therefore, $F^{K^{-1}}_{-1} - F^{K^1}_1$ is increasing in $p_x$ and decreasing in $p_y$, as in Lemma \ref{lem:FdminusFhincreasing}. As a consequence, both Propositions \ref{prop:main_result} and \ref{prop:pyincreasegood} continue to hold for the same reason as in our main model. If $p_x$ increases, we can go from $F^{K^{-1}}_{-1} - F^{K^1}_1 > 0$ to $F^{K^{-1}}_{-1} - F^{K^1}_1<0$. If the likelihood of type $K^{-1}$ is sufficiently large, then this can result in more inefficiencies, exactly as in our model with two types. Similarly, the effect of increasing $p_y$ is also identical to that in our main model.

\subsection{General Signal Structures}
At first glance, it may appear that our result relies heavily on the assumption that the information is binary and that the precision of each signal is symmetric regarding type I and type II errors. However, that is not the case. We have chosen to present our results in the baseline model with a binary signal structure since it makes it considerably easier to elucidate the mechanism clearly. In Appendix \ref{sec:GeneralStructure} we formally demonstrate how our model extends. We generalize our model to a setting in which the verifiable and unverifiable information can come from a continuum. In such an environment, there could be several measures of precision. We propose an order---the spreading order---by comparing information according to how spread out it is around the efficient cutoff (that is, the posterior belief above which it is efficient to act). Loosely speaking, more spread-out information causes  the posterior distributions to be more extreme relative to the efficient belief. Importantly, the spreading order is a strengthening of the Blackwell order.\footnote{Although not exactly the same, the rotation order defined in \citet{johnson2006simple} shares most features with our spreading order.}

We show that a more spread-out verifiable signal can decrease welfare, while a more spread-out unverifiable signal is always welfare increasing. Importantly, the mechanics are identical to those in the binary world: screening the critical types becomes easier with a more spread-out unverifiable signal but harder with a more spread-out verifiable signal.

\section{Applications}\label{sub:examples} 
Before concluding, we wish to consider some  settings---within and outside the formal legal system---to which our model applies. 

First, consider a set of bureaucrats of which some are corrupt, others work with society's interests in mind. Each bureaucrat (the agent) decides whether to approve the expenditure on a certain project, which may be overpriced. Approving expenses means taking a risk because failed, overpriced projects may lead to corruption charges against the bureaucrat. The bureaucrat relies on verifiable information (for example, reports) and unverifiable information (for example, expertise)  to inform himself  whether the project is overpriced and to decide whether  to approve it. The punishment for overpricing depends on the verifiable but not the  unverifiable information. If the bureaucrat is found guilty of corruption, he is sentenced by the court. 

Second,  consider a doctor deciding on the  method for delivering a baby. The doctor relies on some verifiable information (for example, examinations indicating the fetus's position in the womb) and on some unverifiable information (for example, his expertise and tacit knowledge about the fetus). While a C-section is the best method in some cases, choosing an unnecessary C-section risks dire consequences for the mother and the baby. Different doctors value compensation and their own time differently, and C-sections  pay substantially higher and are scheduled. If a C-section leads to complications, and if  no verifiable evidence supports the doctor's choice, he may face legal and administrative consequences. In such a case, if the examiner (a court or hospital administration) concludes that the doctor's interests are not aligned with the patient's, it may wish to punish the doctor by, for example, temporarily revoking his privileges. It applies the reasonable-person standard: to infer the doctor's underlying preferences it compares his behavior to that of a (hypothetical) unbiased doctor.

As a third example, consider a president (the agent) deciding on a foreign policy issue---for instance,  whether to impose sanctions on a country in response to its invasion of a neutral country. This is a risky decision, as the president's electoral chances  might be compromised following negative outcomes. The safe option is to follow standard diplomatic procedures. Different politicians may value the welfare of their constituency differently, especially when weighing it against the wishes of particular special interest groups. The decision is made based on top-secret information (unverifiable information) and  news reports (verifiable information). Voters might hold the president accountable and might want to reelect him only if he values their interests above those of special interests. However, voters have access only to the outcome and the verifiable information.

Finally, consider a CEO of a firm deciding whether  to acquire a smaller firm. The acquisition is risky and may affect the acquiring firm's value and stock price; and the CEO's compensation package with his current and future employers may depend on its outcome. 
When the CEO decides whether to proceed with the acquisition, he relies on verifiable information about the firm to be acquired but also on unverifiable information about the synergy between the firms and about the general outlook of the market. Different CEOs might weigh long-run and short-run outcomes differently. Reviewing a failed acquisition, major shareholders may want to fire a CEO that is interested only in short-term outcomes, while they may wish to retain one that is interested in the long-run development of the firm.

\section{Conclusion}

This paper highlights that it is not merely the amount of information but  its nature that has important welfare consequences. We consider a setting in which an agent decides under uncertainty and may be held liable in court if the decision causes harm. We focus on information of two different natures: that which is verifiable in court and that which is not. We show that increasing the information available to the agent has different consequences depending on its nature.  While increasing the precision of unverifiable information always increases welfare, increasing the precision of verifiable information may reduce it.

Our findings  extend to a variety of settings well beyond legal systems. Whether we consider politicians seeking reelection, CEOs wanting to extend their contracts, or bureaucrats with career concerns, our results apply whenever the principal's ex post evaluation of a risky decision is based only on part of the information available to the agent. The principal has to balance the chilling effect against the desire to deter, and changes in the information structure influence her ability to do so. Our findings are robust to a variety of changes in the assumptions. While details in the timeline or the principal's choice set may differ, the main result remains. The welfare effects of a change in the precision of the information depend on the nature of that information.

The main driver of our result---the tension between deterrence and the chilling effect---has been extensively documented in the legal and management literatures as well as in the popular press.\footnote{See, for instance, \cite{hylton2019economic},  \cite{chalfin2017criminal}, \cite{bernstein2014transparency}, and \cite{bibby1966committee} } Our results show that whether the chilling effect is pronounced enough to outweigh the overall gains from more information is an empirical question. Thus, a natural direction for future research is to empirically quantify the impact of the chilling effect and its interaction with the provision of information of different natures.

\appendix
\singlespacing
\section{Main Results: Proofs}

\subsection{Notation and Cases} 
\label{sub:notation_and_cases}

\paragraph{Cases.} The posterior belief, $\b_{xy}$, depends on the informational environment. We ignore the trivial cases in which signals are irrelevant, either because $\beta_{xy}\leq 1/2~ \forall (x,y)\in \{-1,1\}^2$ or because $\beta_{xy}\geq 1/2~ \forall (x,y)\in \{-1,1\}^2$. What remains are parameter values for which we are in exactly one of the following cases.
\begin{enumerate}
  \item Efficient to act $\Leftrightarrow$ $x=1$;
  \item Efficient to act $\Leftrightarrow$ $y=1$;
  \item Efficient to act $\Leftrightarrow$ $x+y\ge0$;
  \item Efficient to act $\Leftrightarrow$ $x+y=2$.
\end{enumerate}
Case 1 implies that $\pubsigr$ is more informative than $\pvtsigr$. Case 2 implies the reverse. Moreover, a positive realization of the more informative signal is necessary and sufficient to make the project efficient in these cases. Cases 3 and 4 impose no clear ranking between the two types of information. Case 3 implies that $\beta$ is high and that a necessary and sufficient condition for efficiency is that \emph{one} of the signal realizations is positive. Finally, case 4 implies that $\beta$ is low and that a necessary and sufficient condition for efficiency is that \emph{both}  signal realizations are positive. 

\paragraph{Notation.} Let $q^u := a^u(-1,1)$ and $q^b:= a^b(-1,-1)$ be the agent's best responses to $\overline{F}$ and $F(x)$. Notice that if $F(-1) < (>) F^b$, then $q^b =1 (0)$, and if $F(-1) < (>)F^u$, then $q^u = 1(0)$. Let $\eta^u$ and $\eta^b$ be defined by,
\begin{align}
\frac{\g(1-p_y) }{\g(1-p_y) + (1-\g)(1-p_y + p_y \eta^b)}=& \overline{\gamma} \label{Equation: indifference condition for D type on -1,-1}\\
 \frac{\g \eta^u}{\g \eta^u + (1-\g) }=& \overline{\gamma} \label{Equation: indifference condition for H type on -1,1}
\end{align}
If $q^u =1$ then $q^b= \eta^b \implies \gamma_{-1} = \overline{\gamma}$, making the court indifferent between any sentence. If $q^b = 0$ and $a^b(-1,1) =1$, then $q^u= \eta^u \implies \gamma_{-1} = \overline{\gamma}$. 

Finally, let $\overline{W}(\overline{F})$ be the welfare in a welfare-maximizing perfect Bayesian equilibrium conditional on holding the designer choice fixed at $\overline{F}$.

\blemma Define $\of^*:=\arg \max_{\of} \overline W(\of).$ Then, $\of^* \in \{0,F^u,F^b\}$ \elemma

\bprf 
\bclaim\label{Claim: Agent mixing on -1,-1} $q^u = 1 \implies q^b\in \{\eta^b,1\}$ wlog. \eclaim 
\bprf $q^b = 0 \implies \gamma_{-1}= \g> \overline{\gamma}$. Therefore, $F(-1)=0$. Therefore, the biased agent would deviate to play $q^b = 1$. Therefore, $q^b > 0$. Also,  $q^b =1 \implies$ $\gamma_{-1} = \frac{\g (1-p_y)}{\g (1-p_y) + 1-p_y} < \overline{\gamma}$. Therefore, $F(-1) = \of$. Notice that $\of < F^b \implies q^b = 1$. And, if $\of\ge F(-1) > F^b \implies q^b=0 \implies  \gamma_{-1} = \g > \overline{\gamma}$. This would imply that $F(-1) = 0$, a contradiction. Therefore, if $\of > F^b$, the biased type would mix to have $\gamma_{-1} = \overline{\gamma}$---i.e., $q^b = \eta^b$, so that $F(-1) = F^b$.  In the case when $F = F^b$, $q^b \in[ \eta^b,1]$. In this case, $q^b = \eta^b$ is the designer-preferred equilibrium.  \eprf 

\bclaim $ F^u >   F^b$ and $q^b > 0 \implies q^u = 1$. \eclaim 
\bprf $q^b > 0 \implies F(-1) \le F^b <  F^u \implies q^u = 1$. \eprf

\bclaim\label{Claim: Fh > Fd then fine Fh or Fd} If $F^u > F^b$, $\of^* \in \{F^b, F^u\}$. \eclaim 
\bprf First, notice that $F(-1) < F^b \implies q^u = q^b =1.$ 
Instead, with $F(-1) =F^b \implies q^u =1, q^b = \eta^b$, giving us a strict improvement in efficiency. 

If $F(-1) \in (F^b, F^u)$, then $q^u =1$ and $q^b = 0$. But then, $\gamma_{-1} = \g > \overline{\gamma} \implies F(-1) = 0$, a contradiction. Therefore, $F(-1) \notin (F^b, F^u)$ in equilibrium. 

If $F(-1) > F^u$ then $q^u =q^b= 0$. Instead, $F(-1) = F^u$ provides a strict efficiency improvement by having $q^u \in [0, \eta^u], q^b = 0$. The optimal choice is to have $q^u = \eta^u$. $q^u \le \eta^u$ because, otherwise, $\gamma_{-1} > \overline{\gamma}$, and, therefore, $F(-1) = 0$, a contradiction. 
\eprf 

\bclaim\label{Claim: Fh < Fd then fine 0 or Fd} If $F^u < F^b$, $\of^*\in \{0, F^b\}$. \eclaim 
\bprf 
Here, whenever $q^u >0 $, $q^b = 1$. Therefore, either $q^u =q^b =1$, achieved by $\of = 0$, or $q^u = q^b=0$, achieved by $\of = F^b$. Which of the two is optimal depends on whether $\ow(0) > \ow(F^b)$ or vice-versa. It is easy to check that,
\begin{align*}
\ow(0) - \ow(F^b) =& \g [\beta (1-p_x) p_y - (1-\beta) p_x (1-p_y) ] \\
&+ (1-\g) [\beta (1-p_x) (1-p_y)  - (1-\beta)p_x p_y]. 
\end{align*}
Therefore, if $\g$ is sufficiently high, $\overline{F} = 0$; otherwise, $\overline{F} = F^b$. 
\eprf 
Together, the claims imply that $\of^* \in \{0,F^b,F^u\}$. 
\eprf 

\subsection{Proof of Proposition \ref{prop:main_result} and Lemma \ref{lem:FdminusFhincreasing}}
Now we are equipped to present our main comparative static. To this end, let $W^*(\cdot, \cdot, \cdot, \overline{F}; \cdot, \cdot) : = \ow (\of^*)$ denote the optimal equilibrium given $S$ and the given equilibrium. Let $\Delta(p_x,p_y):=F^b-F^u$.

\bprf[Proof of Lemma \ref{lem:FdminusFhincreasing}]
\begin{align*}
F^b =& \frac{1}{1-\b_{-1,-1}} = 1-\frac{\beta}{1-\beta}\frac{1-p_y}{p_y} + \frac{\beta}{1-\beta}\frac{1-p_y}{p_y}\frac{1}{p_x} \quad \\
F^u = &-2 + \frac{1}{1-\b_{-1,1}} = -2 + \frac{-\beta}{1-\beta}\frac{p_y}{1-p_y} + \frac{\beta}{1-\beta}\frac{p_y}{1-p_y}\frac{1}{p_x}\\
\implies \De(p_x,p_y) =& 2 + \frac{\beta}{1-\beta}\frac{1-p_x}{p_x}\left[\frac{1-p_y}{p_y} - \frac{p_y}{1-p_y}\right] 
\end{align*}
The above is increasing in $p_x$ and decreasing in $p_y$. 
\eprf 

\bprf[Proof of Proposition \ref{prop:main_result}]
Fix some $(p_y,\beta)$. At $p_x^*$, $F^b(p_x^*) = F^u(p_x^*)$. Suppose that $p_1 < p_x^* < p_2$. 
Therefore, $F^b(p_1) < F^u(p_1)$ \emph{and} $F^b(p_2) > F^u(p_2)$ by Lemma \ref{lem:FdminusFhincreasing}.

\textbf{Case 1:} $\overline F^*(p_2) = 0$.\footnote{$\overline{F}^*(p)$ denotes $\overline{F}^*$ in the environment with $p_x=p$ ceteris paribus.} 

Hence, $q^b(p_2) = q^u(p_2) = 1$,  By Claim \ref{Claim: Fh > Fd then fine Fh or Fd}, $\overline F^*(p_1) \in \{F^u(p_1), F^b(p_1)\}$. 
Suppose that $F(-1) = F^b(p_1)$. Therefore, $q^b(p_1) = \eta^b$ and $q^u(p_1) = 1$. Notice that \eqref{Equation: indifference condition for D type on -1,-1} features no dependence on $p_1$ and $\eta^b$ is strictly less than $1$.

Let $W_1 := \ow(F^b(p_1))$ and $W_2 := \ow(0)$. 
\begin{align*}
W_i =& \beta\left[ p_i + (1-p_i)[  p_y + (1-\g) (1-p_y) q^b(p_i)] \right] \\
&- (1-\beta) \left[ (1-p_i) + p_i [ (1-p_y) + (1-\g) p_y q^b(p_i) \right]
\end{align*}

Therefore,
\begin{align*}
W_1 -W_2 =& (p_1 -p_2) [ \beta (1-p_y) + (1-\beta)p_y]  \\
&+ (1-\g) \Bigg[ \eta^b \big[ \beta(1-p_1)(1-p_y) - (1-\beta)p_1 p_y\big]\\
&\quad - \big[\beta (1-p_2)(1-p_y) - (1-a) p_2 p_y\big] \Bigg]
\end{align*}

Suppose that for a small $\de > 0$, $p_1 = p_2 - \de$. Then, $$W_1 - W_2 = (1-\g) (1-\eta^b) [ (1-\beta)p_y  p_1 -\beta (1-p_y)(1-p_1)] + o(\de).$$ Since it is inefficient to act on $(-1,-1)$,  $\b_{-1,-1} = \frac{\beta (1-p_y) (1-p_1)}{\beta (1-p_y)(1-p_1) + (1-\beta) p_y p_1} < \frac12$. Equivalently, $(1-\beta) p_y p_1 > \beta (1-p_y)(1-p_1)$. Therefore, $W_1 > W_2$. Lastly, if $\overline{F}^*(p_1) = F^u(p_1)$, then $W_1 > W_2$ for small enough $\delta>0$.
\smallskip

\textbf{Case 2:} $\of^*(p_2) = F^b(p_2)$. 

Therefore, $q^b(p_2) = q^u(p_2) = 0$. Setting $F(-1) = F^u(p_1)$, we have $q^b(p_1) = 0$ \emph{and} $q^u(p_1) = \eta^u > 0$.  Since the only change is that the unbiased type acts on $(-1,1)$ with probability $\eta^u$, the extent of the chilling effect is reduced. Therefore, as before, $W^*(p_1) > W^*(p_2)$ as $\delta \rightarrow 0$. 
\eprf 

\subsection{Proof of Proposition \ref{prop:pyincreasegood}} 
\label{sec:proof_of_proposition_prop:pyincreasegood}

\begin{proof}
We prove the proposition here only for the \emph{interior} of our case. We do so by looking at two types of arguments. Applying these arguments in various combinations is, in fact, sufficient to prove all other cases and the transition from one case to another. We do that in Appendix \ref{sec:general_proof_of_proposition_2}.

The court can observe $x$, the realization of $\pubsigr$. Thus, we can look at the cases separately and provide an argument for each.

\begin{description}
  \item[Argument 1 ($x=1$).] As long as we remain inside our case, the court provides a free pass ($F(x=1)=0$) on realization $x=1$ for any level of $p_y$. In addition, both types act whenever they see $x=1$ and ignore signal $p_y$ entirely. Thus, any improvement on $p_y$ conditional on a realization $x=1$ does not affect the welfare.
  \item[Argument 2 ($x=-1$).] Compare two environments with $p_y, p_y'$ such that $p'_y>p_y$.    
First, assume that $\overline{F}^\ast=0$ for both levels. Increasing precision does not change $a^b(\cdot)$, but projects implemented by the unbiased agent fail less often. Second, assume that $\overline{F}^\ast=F^u$ for both levels. Then, no agent acts when it is inefficient to act (yet there is a moderate chilling effect: see Table \ref{Table: equilibrium strategy profiles}). Because precision increases, the signal on $(-1,1)$ is stronger and welfare improves. Third, assume that $\overline{F}^\ast=F^b$ for both levels. Since $\eta^b$ decreases in $p_y$, the biased agent's actions on $y$ improve from an efficiency perspective, while the unbiased agent's decisions can only improve by Lemma \ref{lem:FdminusFhincreasing}. Welfare increases. What remains is to show that welfare improves as we move from $\of=0$ to $\of=F^\omega$. A change from $\of=F^0$ to $\of=F^b$ occurs either if $F^b>F^u$ or if $F^b=F^u$. In the former case, both equilibria are available, and the switch occurs because $\overline{W}(F^b)$ overtakes $\overline{W}(0)$, an improvement in welfare. In the latter case, welfare improves because the only behavioral change is that the biased agent selects the inefficient action less often. Finally, a change from $\of=0$ to $\of=F^u$ can occur only at $F^b=F^u$, and, by construction, $\of=F^u$ dominates $\of=F^b$. The proof is complete. \qedhere
\end{description}
\end{proof}

\section{Objective Mens Rea: Characterization and Proofs}\label{Appendix: Objective mens rea}
\paragraph{Equilibrium Characterization.} The court is indifferent if $q=\overline{\gamma}$. If $\overline{F}=F^b>F^u$, the optimal equilibrium implies that $a^u(-1,1)=0$, $a^b(-1,1)=1$ and $a^b(-1,-1)=\eta_1$ with 

$$ \eta_1= \min\left\{\frac{(1-p_y)}{p_y} \frac{(1- \overline{\gamma})  }{\overline{\gamma}},1\right\}.$$

If $\overline{F}=F^b<F^u$, the optimal equilibrium implies that $a^u(-1,1)=1$, $a^b(-1,1)=1$ and $a^b(-1,-1)=\eta_2$ with
$$\eta_2= \min\left\{\frac{(1-p_y)}{p_y} \frac{(1- \overline{\gamma})  }{\overline{\gamma}}\frac{1}{1-\gamma},1\right\}.$$

Recall, that  $F^b-F^u$ does not depend on the court's choice, it is still given by

\[\De(p_x,p_y) = 2 + \frac{\beta}{1-\beta}\frac{1-p_x}{p_x}\left[\frac{1-p_y}{p_y} - \frac{p_y}{1-p_y}\right]\]   

If $F^b>F^u$, any punishment below $F^b$ implies that the biased agent is never deterred from acting. If, in addition, $\overline{F}>F^u$,  the unbiased agent is deterred from acting on $(-1,1)$, which is clearly worse. Thus,
an optimal equilibrium exists for	 either $\overline{F}=0$ or $\overline{F}=F^b$. The court's indifference condition implies $\eta_1$.

If $F^b<F^u$, a punishment above $F^b$ does not improve upon $F^b$, as it would lead to actions only on $(-1,1)$, which, in turn, implies that the court does not punish. Conditional on not facing punishment, the biased type has an incentive to deviate and act on both $(-1,1)$ and $(-1,-1)$, which, in turn, implies that not punishing is suboptimal. 
No punishment yields a better outcome than the optimal equilibrium under $\overline{F}=F^b$. Thus, it is sufficient to consider $\overline{F}=F^b$ only if $F^b<F^u$. The court's indifference condition implies $\eta_2$. 

\paragraph{Proof of Proposition \ref{prop:welfarelarger}.}
The level of $F^b$ is unaffected by the court's objective, and so is the ranking $F^b$ vs $F^u$. It suffices to show that welfare is lower for $\overline{F}=F^b$. For $\overline{F}=0$, welfare is, by construction, identical, and $\overline{F}=0$ is selected only if it improves upon $\overline{F}=F^b$. Similarily, $\overline{F}=F^u$ is selected only if it improves on $\overline{F}=F^b$ in the baseline case and never under the objective mens rea. Thus if equilibria conditional on $\overline{F}=F^b$ are welfare-inferior for one court objective, the optimal equilibrium is welfare-inferior under that objective.

To see that result, observe that action profiles are identical, apart from the biased agent's decision on $(-1,-1)$. If $F^b<F^u$ she chooses $\eta_1>0$ for the court's objective assumed in this section (punishing for acting on wrong information) and $0$ under the court's objective in the baseline model.\footnote{For convenience, we call the court's objective in the baseline case as the ``baseline object'' and the court's objective in this section as the ``alternative objective''. } Since acting is inefficient for the information $(-1,-1)$, the alternative objective is welfare-inferior. If $F^b>F^u$, the agent chooses

$$\eta_2= \max\{\frac{(1-p_y)}{p_y} \frac{(1- \overline{\gamma})  }{\overline{\gamma}}\frac{1}{1-\gamma},1\}> \frac{1-p_y}{p_y} \frac{\gamma- \overline{\gamma}}{\overline{\gamma} }\frac{1}{1-\gamma}= \eta^b.$$
Again, the alternative objective is welfare-inferior.

\paragraph{Proof of Proposition \ref{result_objectivemensrea}.}
The first part follows by using the parameters that are used for the figures. Alternatively, one can use a constructive version similar to that of the proof of Propositions \ref{prop:main_result}. We omit it, as it provides no further insight. We discuss the second part below.

\textbf{When is welfare unambiguously increasing in the precision of $p_y$?}

First, consider $p_y < p_y' < p_x^*$ such that $p_y,p_y' \in \mc Y(\beta,p_x)$. Here, the equilibria from the top row of Table \ref{Table extension: equilibrium strategy profiles} are available. It is easy to check that welfare is continuous and increasing in $p_y$ for each of these equilibria. Therefore, $W^*(\beta,p_x,\cdot,\g)$, which selects the maximum of the welfare generated by the two equilibria, is also continuously increasing on $[\underline{p_y}(\beta,p_x),p_y^*)$. 

Using a similar argument $W^*(\beta,p_x,\cdot,\g)$ is continuously increasing on $(p_y^*,\overline{p_y}(\beta,p_x]$. 
Finally, a switch from $p_y$ to a $p_y'$ such that $p_y' > p_y^*> p_y$ that entails switching from $\overline F=0$ to $\overline F = F^b$ is also welfare improving as it only increases deterrence without having a chilling effect. Therefore, 
the only case we need to consider is the case in which $\overline{F}=F^b$ on both sides of $p_y^\ast$,  and precision increases from $p_y<p^\ast_y$ to $p_y'>p_y^\ast$. In all other cases, welfare increases in $p_y$.

A necessary and sufficient condition for the designer to prefer $\overline{F}=F^b$ over the free pass when $p_y<p_y^\ast$ is $W(p_y)$ is higher under $\overline{F}=F^b$. That is the case when
\footnotesize
\[ \begin{split}
\beta\Big(p_y(1-p_x)+ (1-\gamma)(1-p_x)(1-p_y)\Big)-(1-\beta)\Big(p_x(1-p_y)+(1-\gamma)p_x p_y\Big) &> \\
\beta\Big(p_y(1-p_x)(1-\gamma)+ (1-\gamma)(1-p_x)(1-p_y)\eta_1\Big)-(1-\beta)\Big(p_x(1-p_y)(1-\gamma)+(1-\gamma)p_x p_y\eta_1\Big)
\end{split}\]
\normalsize
which can be simplified to
\begin{equation}\label{eq:Fdoptimal}
  \frac{1-\gamma}{\gamma} (1-\eta_1) >  \underbrace{\frac{\beta p_y (1-p_x) - (1-\beta) p_x(1-p_y)}{(1-\beta)p_x p_y - \beta (1-p_x)(1-p_y)}}_{:=\widehat{\Delta}(p_y)}>0
\end{equation}

where the last inequality follows because\textemdash by assumption\textemdash it is efficient to act when any signal is positive.

Next consider the case in which $\overline{F}=F^b$ and define
\begin{align*}
f_1(p_y) :=& \beta \left[p_x + (1-p_x)(1-\g) [ p_y + (1-p_y) \eta_1 ] \right]\\
& - (1-\beta) \left[ (1-p_x) + p_x (1-\g) [ 1-p_y + p_y \eta_1] \right]\\
f_2(p_y) := & \beta \left[ p_x + (1-p_x) [p_y + (1-p_y) (1-\g) \eta_2 ] \right]\\
&  -(1-\beta) \left[ (1-p_x) + p_x [(1-p_y) + p_y (1-\g) \eta_2].\right]
\end{align*}

Notice that $W^*(p) = f_1(p)$ if $p< p_y^*$ and $W^*(p) = f_2(p)$ if $p_y' \ge p^*_y$. Both $f_1(\cdot)$ and $f_2(\cdot)$ are increasing in $p_y$. Thus, if $f_2(p^*_y) \geq f_1(p^*_y)$, welfare is increasing in $p_y$ also around $p_y^\ast$. Otherwise, it is not.

Formally,  
\begin{align*}
f_2(p^*_y) - f_1(p^*_y) =& \beta (1-p_x) p^*_y \g + \beta (1-p_x) (1-p^*_y)(1-\g)(\eta_2 - \eta_1) \\
& -(1-\beta) p_x (1-p^*_y) \g - (1-\beta) p_x p^*_y (1-\g) (\eta_2 -\eta_1)
\end{align*}
or equivalently
\begin{align*}
f_2(p^*_y) - f_1(p^*_y) =& \g \underbrace{[\beta(1-p_x)p^*_y - (1-\beta) p_x (1-p^*_y)]}_{>0} \\
&- (1-\g)(\eta_2 - \eta_1)\underbrace{ [(1-\beta) p_x p^*_y- \beta (1-p_x) (1-p^*_y)]}_{> 0}
\end{align*}
The signs of the two quantities above follow from the fact that it is efficient to act on $(-1,1)$ and inefficient to act on $(-1,-1)$. Thus, welfare increases around $p_y^\ast$ if and only if

\begin{equation}\label{eq:Deltalargeenough}
  \frac{(1- \gamma)}{\gamma} (\eta_2- \eta_1) \leq \underbrace{\frac{ \beta(1-p_x)p^*_y - (1- \beta) p_x (1-p^*_y)}{ (1- \beta) p_x p^*_y- \beta (1-p_x) (1-p^*_y)}}_{=\widehat{\Delta}(p_y^\ast)}.
\end{equation}

Notice that if condition \eqref{eq:Deltalargeenough} is violated for $p_y^\ast$ it is also optimal to implement $\overline{F}=F^b$ for $p_y^\ast$ because $1-\eta_1 \geq \eta_2-\eta_1$ and hence a violation of \eqref{eq:Deltalargeenough} implies \eqref{eq:Fdoptimal}. Thus, a necessary and sufficient condition for Proposition 2 to hold is that \[(\eta_2-\eta_1) \frac{(1-\gamma)}{\gamma}\leq \widehat{\Delta}(p_y^\ast).\]

Observe that $\widehat{\Delta}(p_y^\ast)$ is independent of the courts threshold belief $\overline{\gamma}$. Moreover,

\[ \eta_2- \eta_1 = \begin{cases}
0  &\text{ if }\overline{\gamma}\leq 1-p_y\\
1- \frac{1-p_y}{p_y} \frac{1- \overline{\gamma}}{\overline{\gamma}} &\text{ if } 1-p_y < \overline{\gamma} < \frac{1-p_y}{1-p_y \gamma} \\
\frac{\gamma}{1- \gamma} \frac{1-p_y}{p_y} \frac{1-\overline{\gamma}}{\overline{\gamma}} &\text{ if } \overline{\gamma}\geq \frac{1-p_y}{1-p_y \gamma}.
\end{cases}\]

Notice further that $\eta_2-\eta_1$ is increasing in $\overline{\gamma}$ if and only if $\overline{\gamma} \in [1-p_y,\frac{1-p_y}{1-p_y \gamma}]$ and therefore its maximum at $\overline{\gamma}=\frac{1-p_y}{1-p_y \gamma}$ where $\eta_2-\eta_1=\gamma$ which implies that $\eta_1-\eta_2 \in [0,\gamma]$.

Thus, independent of $\overline{\gamma}$, \eqref{eq:Deltalargeenough} holds if

\[(1- \gamma) \leq \frac{ \beta(1-p_x)p^*_y - (1- \beta) p_x (1-p^*_y)}{ (1- \beta) p_x p^*_y- \beta (1-p_x) (1-p^*_y)}\]
which can be simplified to
\begin{equation}\label{eq:simplified_cond}
  \frac{1-\beta}{\beta} \frac{p_x}{1-p_x} \leq \frac{1-\gamma(1-p^\ast_y)}{1-p^\ast_y \gamma}.
\end{equation}

Because $p_y>1/2$ the right-hand side of the above larger than 1 which in turn implies that if $\beta>p_x$, then Proposition 2 holds for any $(\gamma,\overline{\gamma})$.

Moreover, $p_y>1/2$ implies that the right-hand side of condition \eqref{eq:simplified_cond} is increasing in $\gamma$ with limit $p_y/(1-p_y)$ as $\gamma \rightarrow 1$ 
\[\frac{1-\beta}{\beta} \frac{p_x}{1-p_x} <\frac{p_y}{1-p_y}\] which holds because we are in the case in which a any positive signal makes it efficient to act. Thus, for any $(\beta, p_x,p_y,\overline{\gamma})$ such that we are in our case of interest, there exists a $\hat{\gamma}<1$ such that if the likelihood that the agent is unbiased $\gamma>\hat{\gamma}$, Proposition 2 holds.

Finally, even if \eqref{eq:simplified_cond} fails, there always is a threshold $\underline{\gamma}^\ast>1-p_y^\ast$ such that Proposition 2 holds if $\overline{\gamma}<\underline{\gamma}^\ast$. The reason is that for $\overline{\gamma}$ low enough $\eta_2-\eta_1=0$.

\section{General Information Structure}\label{sec:GeneralStructure}

Throughout the paper, we have assumed that the verifiable and the unverifiable signals are binary. Here we address how our results carry over to more general information structures.

We proceed as follows. First, we discuss how the model translates from the binary case to the more general case. Second, we provide a notion of being \emph{``spread out''} and discuss how it is a meaningful notion of precision in our context. The spreading order is stronger than the, more familiar, Blackwell order of informativeness. That is, if one distribution is more spread out than another, it is also Blackwell more informative. The reverse, however, does not hold. In the spirit of our results from the baseline model, we then show how a more spread out verifiable information may lead to lower welfare. In comparison, a more spread out unverifiable information always increases welfare. Finally, we show that, the same does not hold for the weaker order of Blackwell informativeness. In particular it is not true that \emph{all} Blackwell better unverifiable information (weakly) improves welfare.

\paragraph{Model.} Let $\stater \in \{-1,1\}$ be a binary random variable describing the state of the world as before.\footnote{Unless stated explicitly, bold faces denote random variables and normal fonts denote their realizations.} In the spirit of our baseline model, there are two signals. The verifiable signal is a $[0,1]$-valued random variable denoted by $\pubsigr$. $\pubsigr$ captures the posterior probability that $\stater = 1$, i.e., $\pubsigr = \prob(\stater = 1 \vert \pubsigr)$. Rather than modelling the unverifiable signal, $\pvtsigr$, separately, we denote the agent's total information, again a $[0,1]$-valued random variable, by $\postr$. That is, given the agent's information, $\postr = \prob(\stater = 1 \vert \pubsigr, \pvtsigr)$. Let $\cdfx(\cdot)$ denote the cummulative distribution function (CDF) of $\pubsigr$ and let $\cdf_{\postgivenpub}(\cdot)$ denote the conditional CDF of $\postr$ given $\pubsigr = x$. With some abuse of notation, we will denote by $\postr_{\vert x}$ a random variable with a CDF $\cdf_{\postr_{\vert x}}$. Finally, let $G_{\postr}(\cdot)$ denote the unconditional CDF of $\postr$.  
Bayes plausibility implies that $\E[\pubsigr]= \E[\postr] = \beta$.  Moreover, $\E[\postr \vert \pubsigr] = \pubsigr$. Let the space of all verifiable signals be denoted by $\mc X$. Since we want to demonstrate the results in the spirit of our main model for richer information structures, let $\mathcal I$, be the space of signal structures $(\pubsigr,\pvtsigr)$ such that $\cdf_{\postr}(\cdot)$ has no mass points. Unless stated otherwise, we assume that $(\pubsigr,\pvtsigr) \in \mc I$.  

First, since $\pubsigr$ is verifiable, the optimal punishment can condition on the realization $x$. That is, we can treat the problem for each realization of $\pubsigr$ separately. Given a punishment $F(x)\in\real_+$, let $\mu^\omega(F(x))$ denote the critical posterior---the posterior above which an agent of type $\omega$ acts. We obtain the following.
\begin{align*}
F(x) = \frac{1}{1-\mu^b(F(x))} =& -2 + \frac{1}{1-\mu^u(F(x))}
\end{align*}

In general, if some punishment implies a critical posterior of $\mu^u$ for the unbiased type. With some abuse of notation, the critical posterior, $\mu^b$, for the biased type is given by, 
\begin{align}
\mu^b(\mu^u) = \frac12 \left[3 - \frac{1}{2\mu^u -1}\right]\label{Equation: mud muh relation}
\end{align}

In the spirit of our main model, for $\omega \in \{u,b\}$, define 
\begin{align*}
F^\omega(\mu) := -2\ind_{\omega = u} + \frac{1}{1-\mu}.
\end{align*}

As in the baseline model, the designer's problem is separable in the realization of $\pubsigr$. That is, the designer can choose a punishment $F(x)$ for each realization of $x\in \supp(\pubsigr)$ to provide incentives to both types. Therefore, as in the baseline model, the designer's problem is to choose a function $F: [0,1]\to \real_+$ to maximize welfare.  

In the baseline model, we demonstrated how the welfare effects of improving the precision of information depend on the nature of the information. While capturing precision through a scalar for each signal was natural in the case of binary signals, there does not seem to be a natural analog of the same in the case of richer signals. However, motivated by our setting, the spread of information (made precise shortly) captures the main ideas in the baseline model. It highlights the difference between verifiable and unverifiable information. 
Recall that, from the perspective of a risk-neutral designer, the decision rule in a binary action space is simple: Act if the project is more likely to be good, $\mu>1/2$, do not act if the project is likely to be bad, $\mu<1/2$.

The spread of a signal around $1/2$ captures the degree of certainty in each of the two cases. An increase in the spread means that the mass attributed to posteriors far from $\mu=1/2$ increases. A simple way to formalize this is to think about ``quantile-preserving'' spreads around a posterior $\hat{\mu}$. See also \citeauthor{johnson2006simple}'s (\citeyear{johnson2006simple}) notion of rotation for a similar concept in a different context.

\begin{definition}[Spread.]\label{Definition: dispersion}
We say that $\pubsigr$ is \emph{``more spread out''} around $\hat \mu$ than $\pubsigr'$ if $G_{\pubsigr}(a) \ge G_{\pubsigr'}(a)$ for all $a \le \hat \mu$ and $\cdf_{\pubsigr}(a) \le \cdf_{\pubsigr'}(a)$ if $a> \hat \mu$. 

We say that $\pvtsigr$ is ``more spread out'' around $\hat \mu$ than $\pvtsigr'$ if $\postr_{\vert x}$ is more spread out around $\hat{\mu}$ than $\postr'_{\vert x}$ for a.e. $x$.\footnote{Here, $\postr'_{\vert x}$ means a random variable $\postr':= \prob(\stater=1 \vert \pubsigr = x, \pvtsigr')$ with a CDF $\cdf_{\postr'_{\vert x}}$.}

\end{definition}

As we are interested in the spread around $1/2$, we shorten notation using $\succeq_{sp}$ to denote more spread out around $\hat{\mu}=1/2$.

\begin{remark} Notice that, while the measure of spread of $\pubsigr$ only considers the distribution of $\pubsigr$, the same for $\pvtsigr$ is defined for each realization of $\pubsigr$. This is motivated by the fact that the designer's problem is separable across the realizations of $\pubsigr$, and therefore, the measure of spread of the private information conditions on these realizations. Moreover, notice that, if $\pvtsigr$ is more spread out around $\hat \mu$ than $\pvtsigr'$, then $\cdf_{\postr}(a) \ge \cdf_{\postr'}(a)$ for all $a \le \hat \mu$, and $\cdf_{\postr}(a) \le \cdf_{\postr'}(a)$ for all $a \ge \hat \mu$. 
\end{remark}
We will now see that, in the spirit of the main results from our baseline model, a more spread out (around $1/2$) $\pubsigr$ can reduce welfare, while a more spread out (around $1/2$) $\pvtsigr$ always increases welfare. 

\begin{proposition}\label{Proposition:disperse x can harm}There exist information structures $S_1:= (\pubsigr',\pvtsigr)$ and $S_2:=(\pubsigr,\pvtsigr)$ such that $\pubsigr' \succeq_{sp} \pubsigr$, and the payoff to the designer under $S_1$ is strictly lower than under $S_2$. 
\end{proposition}
\bprf[Sketch of proof] We will only outline the structure of the proof here and omit some straightforward arguments to conserve space and notation. The arguments are nearly identical to the proof of Proposition \ref{prop:main_result}. 

To this end, suppose that $\pvtsigr$ is a binary random variable as in the baseline case with precision $p_y$.\footnote{As the following discussion will demonstrate, the result is not driven by the binary nature of $\pvtsigr$.} Therefore, for each realization of $x \in \supp(\pubsigr)$, there are two possible posteriors, $\{\underline{\mu}(x),\bar{\mu}(x)\}$. Moreover, for any $x$ we have

\begin{align*}
\De(x):= F^b(\underline{\mu}(x)) - F^u(\bar{\mu}(x)) = 2 + \frac{x}{1-x}\left[\frac{1-p_y}{p_y} - \frac{p_y}{1-p_y}\right]
\end{align*}
As in Lemma \ref{lem:FdminusFhincreasing}, $\De(x)$ is decreasing in $x$ and $x^*$ is the belief such that $\De(x^*)=0$. Suppose that $\beta > x^*$, the threshold belief where $F^b(\underline{\mu}(x)) = F^u(\bar{\mu}(x))$. We will suppress the dependence on $x$ henceforth and just denote these objects by $F^b$ and $F^u$. Fix an $\epsilon>0$ that is sufficiently small. Consider two signals as follows. 

\begin{align*}
\cdf_{\pubsigr}(x)= \begin{cases} 0 & \text{ if } x < x^* \\
\frac{x-x^*}{2\epsilon} & \text{ if } x \in [x^*,x^*+\epsilon]\\
\frac12 & \text{ if } x^*+\epsilon < x < x_1 \\
\frac12 + \frac{x-x_1}{2\epsilon} & \text{ if } x \in [x_1, x_1 + \epsilon]\\
1 & \text{ if } x> x_1 +\epsilon
\end{cases}
\end{align*}
where $x_1$ is such that $\frac12[x^* +  \frac{\epsilon}{2}] + \frac12 [ x_1 + \frac{\epsilon}{2}] = \beta$. 
And, 

\begin{align*}
G_{\pubsigr'}(x)= \begin{cases} 0 & \text{ if } x < x^*-\epsilon \\
\frac{x-x^*+\epsilon}{2\epsilon} & \text{ if } x \in [x^*-\epsilon,x^*]\\
\frac12 & \text{ if } x^*< x < x_2 \\
\frac12 + \frac{x-x_2}{2\epsilon} & \text{ if } x \in [x_2, x_2 + \epsilon]\\
1 & \text{ if } x> x_2 +\epsilon
\end{cases}
\end{align*}
where $x_2$ is such that $\frac12[x^- \frac{\epsilon}{2}] + \frac12 [ x_2 + \frac{\epsilon}{2}] = \beta$. 

Notice that $x_2>x_1$, and hence, $\pubsigr' \succeq_{sp}\pubsigr$ if $\beta$ such that $x_1 > \frac12$. Suppose that $\beta$ and $p_y$ are such that the following restrictions hold. 
\begin{enumerate}
\item $\bar\mu(x^*-\epsilon) > \frac12 > x^* + \epsilon$. Therefore, $\underline\mu(x^*+\epsilon) < \frac12$. This implies that it is efficient to act on $\bar \mu(x)$ for all $x \in \supp(\pubsigr)$ as well as for all $x\in \supp(\pubsigr')$.  
\item $\underline\mu(x_1) > \frac12$. Notice that $x_2 > x_1$. Therefore, this assumption implies that it is efficient to whenever $\pubsigr\ge x_1$ or $\pubsigr'\ge x_2$. 
\end{enumerate}

The reason why the welfare with $(\pubsigr',\pvtsigr)$ is lower than with $(\pubsigr,\pvtsigr)$ is nearly identical to the baseline model. In the case of $\pubsigr'$, with probability $\frac12$, the posterior is above $x^*$. In this case, $F^b(\underline{\mu}(x)) > F^u(\bar{\mu}(x))$. The designer cannot induce the unbiased agent to act on $\bar\mu(x)$ while preventing the biased agent from acting on $\underline\mu(x)$. Thus, for a sufficiently high $\g$, the optimal equilibrium would entail $F(x) = 0$ and the equilibrium would be as in Table \ref{Table:Fh less than Fd and Fbar =0}. The biased agent acts with probability $1$ for all the realizations. The unbiased agent acts whenever it is efficient to do so. On the other hand, in the case of $\pubsigr$, the posterior is always above $x^*$ and the equilibrium would be as in Table  \ref{Table:Fh greater than Fd and Fbar =Fd}. Here, as in the baseline model, the biased agent acts with an interior probability, $\eta^b$, on $\underline\mu(x)$, while the unbiased agent acts with probability $1$ whenever it is efficient to do so.
Recall that the equilibrium in Table \ref{Table:Fh less than Fd and Fbar =0} yields a strictly lower welfare than the one in Table \ref{Table:Fh greater than Fd and Fbar =Fd}. Therefore, replicating the argument in Proposition \ref{prop:main_result}, for a small $\epsilon$, the reduction in welfare follows, i.e., welfare under $(\pubsigr,\pvtsigr)$ is strictly larger than that under $(\pubsigr',\pvtsigr)$. 
\eprf 

We now turn to the remaining point: To show is that a more spread out $\pvtsigr$ implies an unambiguous increase in welfare as in Proposition \ref{prop:pyincreasegood}. To this end, fix a verifiable signal, $\pubsigr$. Let 
$$\mc Y^{\pubsigr} := \{\pvtsigr: \forall x \in \supp(\pubsigr), \quad  \cdf_{\postr\vert x}(\cdot) \text{ has no atoms.} \}.$$

Let $F(\cdot)$ be some punishment function. Let $\{\mu^u(F(x)), \mu^b(\mu^u(F(x)))\}$ be the posteriors that $F(x)$ makes indifferent for type $u$ and $b$ respectively. Henceforth, we will simply denote them by $\mu^b$ and $\mu^u$. Notice that, without loss, $\mu^b\le \frac12 \le \mu^u$ with at least one inequality being strict. The biased type acts whenever $\mu \ge \mu^b$ while the unbiased type acts whenever $\mu \ge \mu^u$. Therefore, we can define the welfare given a signal $\pvtsigr$ (suppressing the dependence on $\pubsigr$, and its realization, $x$) to be
\begin{align*}
\Pi(\pvtsigr) := (1-\g) \int_{\mu^b}^1 (2\mu-1) \dd \cdf_{\postgivenpub}(\mu) + \g \int_{\mu^u}^1 (2\mu-1) \dd \cdf_{\postgivenpub}(\mu).
\end{align*}

\begin{proposition} \label{Proposition: better Y harms welfare, general structure}Consider $\pubsigr \in \mc X$, and $\pvtsigr, \pvtsigr' \in \mc Y^{\pubsigr}$ such that $\pvtsigr \succeq_{sp} \pvtsigr'$.  Then $\Pi(\pvtsigr') \ge \Pi(\pvtsigr)$. \end{proposition}

\bprf 
We want to establish the following. For all $x \in \supp(\pubsigr)$,  
\begin{align*}
\Pi(\pvtsigr) - \Pi(\pvtsigr') =&~ (1-\g) \int_{\mu^b}^1 (2\mu-1) \dd (\cdf_{\postgivenpub}-\cdf_{\postr'_{\vert x}})(\mu) \\
&+ \g \int_{\mu^u}^1 (2\mu-1) \dd  (\cdf_{\postgivenpub}-\cdf_{\postr'_{\vert x}})(\mu)\\ \ge&~ 0.
\end{align*}
Let $I_1:= \int_{\mu^b}^1 (2\mu-1) \dd (\cdf_{\postgivenpub}-\cdf_{\postr'_{\vert x}})(\mu)$. Define $h(\mu,\mu_1):= (2\mu-1) \ind_{\mu \ge \mu^1} + (2\mu^1 -1) \ind_{\mu < \mu^1}$. Notice that $h(\cdot,\mu^1)$ is an increasing and a convex function in its first argument. Moreover, since $\pvtsigr$ is more spread out around $\frac12$ than $\pvtsigr'$, $(\cdf_{\postgivenpub}-\cdf_{\postr'_{\vert x}})(\cdot)$ changes sign exactly once at $\frac12$. Therefore, by Theorem 3.A.44 of \cite{shaked2007stochastic}, $\postr_{\vert x} \ge_{cx} \postr'_{\vert x}$, where $\ge_{cx}$ denotes the convex order.\footnote{Given two random variables, $Z_1,Z_2$, we say that $Z_1 \ge_{cx} Z_2$ if $\E [\phi(Z_1)]\ge \E[\phi(z_2)]$ for all convex functions $\phi: \real\to \real$.} Therefore, 
\begin{align*}
&\int_0^1 h(\mu,\mu^b) \dd (\cdf_{\postgivenpub}-\cdf_{\postr'_{\vert x}})(\mu) \ge 0 \\
\implies & h(\mu^b,\mu^b) (\cdf_{\postgivenpub}-\cdf_{\postr'_{\vert x}})(\mu^b) + \int_{\mu^b}^1 (2\mu-1) \dd (\cdf_{\postgivenpub}-\cdf_{\postr'_{\vert x}})(\mu) \ge 0
\end{align*}
Since $\mu^b < \frac12$, $h(\mu^b,\mu^b) <0$ and $(\cdf_{\postgivenpub}-\cdf_{\postr'_{\vert x}})(\mu^b) \ge 0$. Therefore, $$I_1 = \int_{\mu^b}^1 (2\mu-1) \dd (\cdf_{\postgivenpub}-\cdf_{\postr'_{\vert x}})(\mu) \ge 0.$$ 
Similarly,  
\begin{align*}
&\int_0^1 h(\mu,\mu^u) \dd (\cdf_{\postgivenpub}-\cdf_{\postr'_{\vert x}})(\mu) \ge 0 \\
\implies & h(\mu^u,\mu^u) (\cdf_{\postgivenpub}-\cdf_{\postr'_{\vert x}})(\mu^u) + \int_{\mu^u}^1 (2\mu-1) \dd (\cdf_{\postgivenpub}-\cdf_{\postr'_{\vert x}})(\mu) \ge 0
\end{align*}

Since $\mu^u > \frac12$, $h(\mu^u,\mu^u) >0$ and $(\cdf_{\postgivenpub}-\cdf_{\postr'_{\vert x}})(\mu^b) \le 0$. Therefore, $$I_2 = \int_{\mu^u}^1 (2\mu-1) \dd (\cdf_{\postgivenpub}-\cdf_{\postr'_{\vert x}})(\mu) \ge 0.$$ 
Therefore, $\Pi(\pvtsigr) -\Pi(\pvtsigr') = (1-\g) I_1 + \g I_2 \ge 0$. 
\eprf 
 
\begin{remark} Notice that Proposition \ref{Proposition: better Y harms welfare, general structure} establishes a ranking for an arbitrary punishment for a given realization $x$. Since $\Pi(\pvtsigr) \ge \Pi(\pvtsigr')$ for all $F(x)$, the welfare ranking analogous to Proposition \ref{prop:pyincreasegood} follows by taking the supremum over $F(x)$ on both sides for each $x$. \end{remark}

\paragraph{Intuition.} Recall the intuition from the baseline (binary) model. Increasing the verifiable information may lead to a qualitative change in the type of equilibria the designer can support. As $p_x$ increases, it may not be possible to effectively separate the critical type-information pairs: the information $(-1,-1)$ for the biased agent and the information $(-1,1)$ for the unbiased agent. As $p_x$ increases, there ceases to be a punishment that deters the former yet encourages the latter. The same mechanism drives Proposition \ref{Proposition:disperse x can harm} for a more general information structure. As $\pubsigr$ becomes more spread out, a higher mass on posteriors prevents the designer from separating the critical type-information pairs. The welfare reduction follows. 

For the unverifiable information, the intuition follows, again, by considering the binary model. In that model, for any given realization of $\pubsigr$, there are $2$ realizations $\pvtsigr$: one good, one bad. As we want to--- in our baseline case at least---motivate the unbiased agent to act after a good $\pvtsigr$ realization, and deter the biased agent from acting after a bad $\pvtsigr$ realization, the posterior after the good (bad) signal lies above (below) $1/2$. As $p_y$ increases, the distance between the two critical posteriors increases. Therefore it becomes easier to separate the two. That exact logic carries over to the general setting. With more spread out signals, there is more mass on the ``easier'' cases which facilitates separation. The result follows.

\paragraph{Informativeness.} At first glance, it may seem surprising that other (and perhaps more familiar) notions of ``better information'' such as e.g., Blackwell informativeness do not deliver our result. The reason is that the Blackwell informativeness order is too weak. Hence, under the Blackwell order, better unverifiable information may also reduce welfare. The spread order strengthens the Blackwell order adapting it to our specific decision problem.\footnote{Blackwell informativeness is defined by a mean-preserving spread instead, that is, if $F$ and $G$ have the same mean, $F$ is a mean preserving spread of $G$ if $\int_{-\infty}^x F(s) ds\ge \int_{-\infty}^x G(s)ds$ for any $x$ strictly inside the joint support. It is straightforward to see, that the spread order satisfies that property. However, not every mean preserving spread is more spread out. Below we discuss such an example.} Because the spread order is stronger than the Blackwell order, the possibility result in Proposition \ref{Proposition:disperse x can harm} holds. However, Proposition \ref{Proposition: better Y harms welfare, general structure} need not hold under improvements according to the Blackwell order. We show this below.

To this end, as before, we say that $\pvtsigr$ \emph{is more informative than} $\pvtsigr'$ if $\postrx$ is Blackwell more informative than $\postrx'$ for all $x\in\supp(\pubsigr)$. 

\begin{figure}[h!]\centering \caption{Posterior beliefs with $\pvtsigr$ and $\pvtsigr'$}
  \includegraphics[scale=1.1]{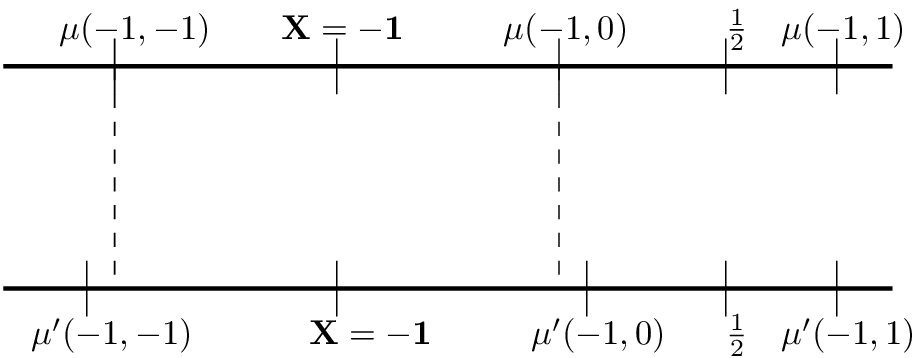}\label{fig:blackwell Y}
\end{figure}

To see how a more informative $\pvtsigr$ can harm welfare, suppose that $\pubsigr$ is a binary random variable as in our baseline model, and, suppose that $\vert \supp(\pvtsigr)\vert = \vert \supp(\pvtsigr')\vert =3$. For the sake of concreteness, suppose that $\supp(\pubsigr) = \{-1,1\}$ and $\supp(\pvtsigr) = \supp(\pvtsigr') = \{-1,0,1\}$. With some abuse of notation, we denote by $\mu(x,y)$ to mean $\postr$ after observing $\pubsigr =x $ and $\pvtsigr = y$. Similarly, $\mu'(x,y)$ denotes $\postr'$ after observing $\pubsigr =x $ and $\pvtsigr' = y$. Suppose that the following holds: Let the signals, $\pubsigr, \pvtsigr, \pvtsigr'$ be such that $\mu(1,\cdot),\mu'(1,\cdot) > \frac12$. Moreover, suppose that the following holds: 
\begin{enumerate}
\item $\postr(1,\cdot) > \frac12, \postr'(1,\cdot) > \frac12$. That is, it is efficient to act whenever $\pubsigr = 1$ regardless of the unverifiable information. 
\item \begin{enumerate}[label=(\roman*)]
\item $\postr(-1,-1) < \postr(-1,0) < \frac12 < \postr(-1,1)$ and, 
\item $\postr'(-1,-1) < \postr'(-1,0) < \frac12 < \postr'(-1,1)$.
\end{enumerate}
Together, these imply that it is efficient to act on $(-1,1)$ for both $(\pubsigr, \pvtsigr)$ and $(\pubsigr,\pvtsigr')$, and inefficient to act on $(-1,-1)$ and $(-1,0)$. 
\item Finally, suppose that $\mu'(-1,0) = \mu(-1,0) + \epsilon_1$ and $\mu'(-1,-1) = \mu(-1,-1) - \epsilon_2$ and $\mu'(-1,1) = \mu(-1,1)$ for a small $\epsilon_1, \epsilon_2 > 0$. 
\end{enumerate}
Figure \ref{fig:blackwell Y} illustrates the posteriors when $\pubsigr = -1$. 

By construction, $\pvtsigr'$ is Blackwell more informative than $\pvtsigr$ as $\pvtsigr'$ is a mean-preserving spread of $\pvtsigr$. Finally, to see how this can, potentially, reduce welfare, the designer wishes to induce the unbiased agent to act on $(-1,1)$ while preventing the biased agent from acting on $(-1,0)$. If, say, $\mu^b(\mu(-1,1)) = \mu(-1,0)$ (recall Equation \ref{Equation: mud muh relation}), then this is feasible. However, when $\mu(-1,0)$ increases to $\mu'(-1,0)$ with $\pvtsigr'(-1,0)$, then the designer can no longer have the unbiased agent act on $(-1,1)$ with positive probability while preventing the biased agent from acting on $(-1,0)$ with probability $1$. The welfare reduction follows due to identical reasoning as in the baseline model. 

We conclude with a brief discussion about information structures in light of these results. Blackwell informativeness has indeed too many degrees of freedom. It cannot determine whether an improvement in either type of information, verifiable and unverifiable, increases or decreases welfare without further qualifications. When two experiments are ranked according to the Blackwell ranking, we obtain a ranking across \emph{arbitrary} decision problems. Instead, our problem has more structure and allows a stronger definition of  `better information.' 
The central friction of our model is that a biased agent may act even when he believes that the project may fail. Likewise, the unbiased agent may refuse to act even when he believes that the project succeeds. More spread out information implies overall more convinced agents. Our finding shows that more spread out unverifiable information is beneficial\textemdash the fear (confidence) of the biased (unbiased) increases. The same does not hold for verifiable information\textemdash being more spread out. Here, it may lead to less trust in unverifiable but valuable signals causing welfare losses even if the punishment tries to account for it.

\section{Cases not discussed in the main text} 
\label{sec:online_appendix}

As mentioned in the main text, we have the following four cases depending on where it is interim efficient to act. 
\begin{enumerate}[label= Case \arabic*., align=left]
\item Efficient to act iff $x=1$. 
\item Efficient to act iff $y=1$.
\item Efficient to act iff $x+y \ge 0$. 
\item Efficient to act iff $x+y=2$. 
\end{enumerate}
In the main text, we analyzed case 3 where it is efficient to act iff either $x$ or $y$ is $1$. We will refer to this case as the baseline case henceforth. 
Case 1 is straightforward, as mentioned in the main text, by setting $F(-1)$ to be very large and $F(1) = 0$. We now analyze the remaining 2 cases.

\subsection{Efficient to Act iff $y=1$.}\label{y1}

In this case, the goal is to deter the $b$ type from acting on $(-1,-1)$ and $(1,-1)$ while incentivizing the $u$ type from acting on $(-1,1)$ and $(1,1)$. 
An optimal policy trades off between different costs just like in the baseline case. The main difference in this case is that we have to choose $F(-1)$ and $F(1)$. Recall that in the baseline case $F(1)$ was $0$ as acting on $x=1$ was efficient. 

The key idea in this case is that by setting $\bar F$ high enough, we can essentially treat the analysis of $x=1$ separately from when $x=-1$. In fact, on each of these, the reasoning that guides us to $\bar F^*(-1)$ and $\bar F^*(1)$ is identical to those from the baseline case. We state the claims for this environment below. The proofs are identical and hence we have chosen to skip them. 

Let $F^u(1,1)$ denote the largest punishment up to which the unbiased type will act on $(1,1)$ and $F^b(1,-1)$ denote the smallest punishment necessary to deter the biased type from acting on $(1,-1)$.

\bclaim\label{Claim: online appendix if Fh0 > Fd0 then Fstar0 in Fd0 Fh0. And the other case}If $F^u(-1,1) > F^b(-1,-1)$ then $\bar F^*(-1) \in \{ F^b(-1,-1), F^u(-1,1)\}$. If $F^u(-1,1) < F^b(-1,-1)$ then $\bar F^*(-1) \in \{0, F^b(-1,-1)\}$. \eclaim 

\bclaim\label{Claim: online appendix if Fh1 > Fd1 then Fstar0 in Fd1 Fh1. And the other case}If $F^u(1,1) > F^b(1,-1)$ then $\bar F^*(1) \in \{ F^b(1,-1), F^u(1,1)\}$. If $F^u(1,1) < F^b(1,-1)$ then $\bar F^*(1) \in \{0, F^b(1,-1)\}$. \eclaim

As in the baseline case, for the main comparative static, the difference between $F^b(-1,-1)$ and $F^u(-1,1)$, as well as the difference between $F^b(1,-1)$ and $F^u(1,1)$ matters in determining the optimal fine. Define, 
\begin{align*}
\Delta^1(p_x,p_y) :=& F^b(1,-1) - F^u(1,1) = 2 + \frac{\beta}{1-\beta}\frac{p_x}{1-p_x}\left[ \frac{1-p_y}{p_y} - \frac{p_y}{1-p_y}\right]\\
\Delta^{-1}(p_x,p_y) :=& F^b(-1,-1) - F^u(-1,1) = 2 + \frac{\beta}{1-\beta}\frac{1-p_x}{p_x}\left[ \frac{1-p_y}{p_y} - \frac{p_y}{1-p_y}\right]
\end{align*}

Below we state a straightforward result, exactly as in Lemma \ref{lem:FdminusFhincreasing} from the main text for the baseline case. 
\blemma \label{Lemma: comparative static driver y=1 case}$\Delta^1(p_x,p_y)$ is decreasing in $p_x, p_y$. $\Delta^{-1}(p_x,p_y)$ is increasing in $p_x$ and decreasing in $p_y$. Moreover, $\Delta^{-1}(p_x, p_y) > \Delta^1(p_x,p_y)$ for all $p_x, p_y \in (\frac12, 1)$. \elemma

Recall that we obtained Proposition \ref{prop:main_result} thanks to the following observation: Start with a $p_x$ such that $\Delta^{-1}(p_x,p_y) < 0$ but is close to $0$. Then, it is possible to have $a^b(-1,-1) = \eta^b < 1$ \emph{and} $a^u(-1,1) = 1$. However, a slight increase from $p_x$ to $p_x' > p_x$, we can have $\Delta^{-1}(p_x',p_y) > 0$. In this case, we can no longer have an equilibrium where $a^u(-1,1) = 1$ \emph{and} $a^b(-1,-1) < 1$. In particular, if $\g$ is sufficiently high we set $\bar F^*(-1) = 0$ and obtain $a^u(-1,1) = a^b(-1,-1) =1$. That is, the unbiased type acts on $(-1,1)$ but the price we pay is that the biased type acts with probability $1$ on $(-1,-1)$. 

Notice that this reasoning is identical, when $x=-1$, in the case where $y$ is pivotal. Also, if $\Delta^1(p_x, p_y) < 0$, then we can have $a^b(1,-1) = \eta^b$ \emph{and} $a^u(1,1) = 1$. Moreover, the crucial point to note is that $\Delta^1(p_x, p_y)$ is decreasing in $p_x$, and is smaller than $\Delta^{-1}(p_x,p_y)$. Therefore, if $\Delta^{-1}(p_x, p_y) < 0$ then $\Delta^1(p_x,p_y) < 0$. And, for any $p_x' > p_x$, $\Delta^1(p_x',p_y) < 0$. As a consequence, if we have a critical belief $p_x^*$, i.e. $F^b(-1,-1) = F^u(-1,1)$, then $\Delta^1(p_x^*,p_y) < 0$, and will continue to be so in a neighbourhood of $p_x^*$. 

Therefore, replicating the construction as in the baseline case, we can obtain a similar result as in Proposition \ref{prop:main_result} and \ref{prop:pyincreasegood} in this environment as well. That is, there exist a set of parameters where increasing $p_x$ can reduce welfare but increasing $p_y$ can never harm welfare. We state them formally below. 

\begin{proposition} \label{Proposition: online appendix both comparative statics}There are (non knife-edge) environments, $(p_x, p_x', p_y,\g,\beta)$ , such that $p_x > p_x'$ and $W^*(p_x') > W^*(p_x)$. Moreover, for all environments $(p_x, p_y, \g,\beta)$  welfare $W^*(p_y)$ is non-decreasing. \end{proposition}

\subsection{Efficient to Act iff $x=y=1$}
First of all, in this case, we can set $F$ to be larger than $F^b(-1,1)$ and convict on $x=-1$. This way, we ensure that no type acts on $x=-1$. Therefore, what remains is the case when $x=1$. Here, we want to have $a^u(1,1) = 1$ and $a^b(1,-1) = 0$. Unsurprisingly, the possibility of this depends on how $F^u(1,1)$ and $F^b(1,-1)$ are ranked. Lastly, since $\Delta^1(p_x, p_y)$ is decreasing in $p_x$, increasing $p_x$ cannot reduce welfare in this case. 

\subsection{General Proof of Proposition 2} 
\label{sec:general_proof_of_proposition_2}
\paragraph{First step: Inside each case.} The main observation is that the cases $x=1$ and $x=-1$ can be addressed separately since the court can condition on the realization of $\pubsigr$ and by Lemma \ref{Lemma: comparative static driver y=1 case} above $\Delta^1(p_x,p_y)$ and $\Delta^{-1}(p_x,p_y)$ decrease in $p_y$.

We restate below the four cases mentioned earlier in this appendix and the main text.
\begin{enumerate}[label = Case \arabic*]
  \item \textbf{It is efficient to act iff $z=1$}. In this case, Argument 1 of Appendix \ref{sec:proof_of_proposition_prop:pyincreasegood} applies for both $z=1$ and $z=-1$
  \item \textbf{It is efficient to act iff $y=1$.} For $z=-1$ this case is identical to the baseline case. Argument 2 of Appendix \ref{sec:proof_of_proposition_prop:pyincreasegood} applies. Also conditional on $z=1$ the situation is as in the baseline case for $z=-1$. Since $\Delta^{1}(p_x,p_y)$ decreases in $p_y$, Argument 2 of Appendix \ref{sec:proof_of_proposition_prop:pyincreasegood} applies directly.
  \item \textbf{It is efficient to act iff $z+y\geq 0$.} This is the baseline case. We showed it in Appendix \ref{sec:proof_of_proposition_prop:pyincreasegood}.
  \item \textbf{It is efficient to act iff $z+y=2$.} The case for $z=1$ is as in the previous case and Argument 2 of Appendix \ref{sec:proof_of_proposition_prop:pyincreasegood} applies. For $z=-1$ Argument 1 of Appendix \ref{sec:proof_of_proposition_prop:pyincreasegood} applies
\end{enumerate}
\paragraph{Second Step: Accross cases.} As we keep $(p_x,\gamma,\beta)$ fixed and increase $p_y$, we can move across cases. In particular, the following relation holds.
\begin{itemize}
\item Case 2 is absorbing---any increase in $p_y$ keeps us in this case. 
\item Case 3 can only transition to case 2. 
\item Case 4 can go to case 2 directly or through case 3. 
\item Case 1 can go through case 3 or case 4.
\item In knife-edge cases, a direct transition from case 1 to case 2 is possible. 
\end{itemize} 
We show that $W^*(p_x,p_y,\g,\beta)$ is continuous at the boundaries and thus, by the first step, welfare improves.  \begin{description}
  \item[From 1 to 3.] Take $\hat{p}_y$ such that  $\mathbb{P}(\theta=1|\pubsigr=-1,\pvtsigr=1)=1/2$. Then, for any $p_y<\hat{p}_y$, we are in case 1 and for any $p_y>\hat{p}_y$ we are in case 3. For $\hat{p}_y$, full deterrence of the biased type without any chilling effect is possible by the punishment $F^b$ conditional on $x=-1$ since the unbiased type (and the designer) are indifferent between taking an action or not taking an action. Yet, $F^b$ is also feasible. Thus, the transition is continuous.
  \item[From 1 to 4.] Take  $\hat{p}_y$ such that  $\mathbb{P}(\theta=1|\pubsigr=1,\pvtsigr=-1)=1/2$. Then, the designer is indifferent between everyone acting on $(-1,1)$, no one acting on it, or only the biased type acting on it. It is feasible by setting $F(-1)=1$ for example. It covers all the potential action profiles in case 4. Thus, welfare is continuous at the boundary.
  \item[From 3 to 2.] Analogous to the case from 1 to 4.
  \item[From 4 to 2.] Analogous to the case from 1 to 3. 
\end{description}

\section{Extensions Not Discussed in the Main Text}
\subsection{Asymmetric Precision}
\label{sub:asy} 
In the baseline model, we assumed that the precision of a signal, $\pubsigr$ or $\pvtsigr$, is independent of the state. That is, $p_x$ is the probability that $\pubsigr$ matches the state regardless of the actual realization of the state. We now relax this. Let $p_x^i:= \mathbb P(\pubsigr = \theta \vert \theta = i)$ and $p_y^i$ analogously. Straightforward calculations show that $$F^b - F^u = -2 + \frac{\beta}{1-\beta}\frac{1-p_x^1}{p_x^{-1}}\left[ \frac{1-p_y^1}{p_y^{-1}} - \frac{p_y^1}{1-p_y^{-1}}\right].$$
Therefore, $F^b-F^u$ is increasing in $p_x^1, p_x^{-1}$ and decreasing in $p_y^1,p_y^{-1}$.\footnote{To be more precise, the said monotonicity holds when, as in our main model, we assume that $p_x^1,p_x^{-1}, p_y^1,p_y^{-1}\ge \frac12$.} The comparative statics follow from this monotonicity. 

\subsection{Conditionally Dependent Signals} 
\label{sub:CondDep}
In the baseline model, we assumed that signals $\pubsigr$ and $\pvtsigr$ are conditionally independent. Relaxing this assumption is not straightforward. We want to retain a structure that allows us to perform comparative statics wherein we improve the precision of one signal while keeping the precision of the other signal constant. Here we provide one specific example. Suppose that $\pubsigr$ (the verifiable information) is a binary signal which has a precision of $p_x$. Moreover, $\pvtsigr$ is another binary signal that is equal to $\pubsigr$ with probability $\rho$. More precisely, conditional on the state $\theta$, $\pvtsigr$ has the following distribution: 
\begin{align*}
\pvtsigr = \begin{cases} 
\pubsigr & \text{ w.p.} \rho \\
\theta & \text{ w.p. } p_y (1-\rho)\\
-\theta & \text{ w.p. } (1-p_y) (1-\rho)
\end{cases}
\end{align*}

In this case, we have, 
\begin{align*}
F^b - F^u = -2 + \frac{\beta}{1-\beta}\frac{1-p_x^1}{p_x^{-1}}\left[\underbrace{\frac{1-(1-\rho) p_y}{\rho + p_y (1-\rho)} - \frac{p_y}{1-p_y}}_{A}\right]
\end{align*}
It is easy to check that $A \le 0$ for any $\rho \in [0,1]$. Therefore, as in Lemma 2, $F^b-F^u$ is increasing in $p_x$ and decreasing in $p_y$. The comparative statics follow from this monotonicity.

\bibliography{bib}

\end{document}